\documentclass[prd,showpacs,floatfix]{revtex4}
\usepackage[dvips]{graphicx,psfrag}
\usepackage{amssymb}
\input epsf
\newcommand*{\be}{\begin{equation}}
\newcommand*{\ee}{\end{equation}}
\newcommand*{\bea}{\begin{eqnarray}}
\newcommand*{\eea}{\end{eqnarray}}
\newcommand*{\lb}{\label}

\begin{document}
\title{Transient Accelerated Expansion and Double Quintessence}
\author{David Blais}
\affiliation{Laboratoire de Math\'ematiques et Physique Th\'eorique, UMR 6083 CNRS,\\
Universit\'e de Tours, Parc de Grandmont, 37200 Tours, France and\\
Laboratoire de Physique Math\'ematique et Th\'eorique, UMR 5825 CNRS,\\
Universit\'e de Montpellier II, 34095 Montpellier, France.}
\author{David Polarski}
\affiliation{Laboratoire de Physique Math\'ematique et Th\'eorique, UMR 5825 CNRS,\\
Universit\'e de Montpellier II, 34095 Montpellier, France.}

\date{\today}

\begin{abstract}
We consider Double Quintessence models for which the Dark Energy sector consists of two 
coupled scalar fields. We study in particular the possibility to have a transient 
acceleration in these models. In both Double Quintessence models studied here, it is shown that if 
acceleration occurs, it is necessarily transient. We consider also the possibility to 
have transient acceleration in two one-field models, the Albrecht-Skordis model and the 
pure exponential. 
Using separate conservative constraints (marginalizing over the other parameters) on the effective 
equation of state $w_{eff}$, the relative density of the Dark Energy $\Omega_{Q,0}$ and the 
present age of the universe, we construct scenarios with a transient acceleration that 
has already ended at the present time, and even with no acceleration at all, but a less 
conservative analysis using the CMB data rules out the last possibility.
The scenario with a transient acceleration ended by today, can be implemented for the range 
of cosmological parameters $\Omega_{m,0}\gtrsim 0.35$ and $h\lesssim 0.68$.  
%
%
\end{abstract}
\maketitle

\section{Introduction}
The release of type Ia Supernovae data independently by two groups, the Supernovae Cosmology 
Project and the High-z Survey Project \cite{SN1}, confirmed in more recent work 
\cite{SN2}, indicating that our Universe might be presently 
accelerating, has profound implications on the current paradigm in cosmology. It was found that, 
assuming flatness, the best-fit Universe with a cosmological constant $\Lambda$ is given by the 
set of cosmological relative densities $(\Omega_{\Lambda,0},\Omega_{m,0})=(0.72, 0.28)$. 
Here, $m$ stands for usual pressureless matter including (cold) dark matter and baryonic matter.
These data are also in surprising agreement with the location of the first acoustic 
(Doppler) peak of the Cosmological Microwave Background temperature anisotropy multipoles. 
If these data are confirmed in the future, they imply a radical departure from usual textbooks 
Friedman-Lema\^{\i}tre-Robertson-Walker (FLRW) cosmology \cite{SSPR}. Indeed, a perfect isotropic 
fluid cannot lead to accelerated expansion unless it has a sufficiently negative pressure while 
the data seem to imply that such a kind of smoothly distributed matter, called Dark Energy, 
constitute about two thirds of the whole matter budget of our Universe. 

Obviously, a pure comological constant $\Lambda$ could be responsible for this acceleration. However, 
its amplitude has to be exceedingly small, about $123$ orders of magnitude too small in order to be 
explained in a ``natural'' way. While this possibility is actually in good agreement with observations 
and attractive as it seems to make many pieces of our present understanding of structure 
formation fit into a consistent picture, in view of the above mentioned theoretical problems research 
has focused on other, more elaborate, Dark Energy candidates that would mimic a pure cosmological 
constant.

A $\Lambda$-term is equivalent to a perfect isotropic fluid with constant equation of state 
$p_{\Lambda}= -\rho_{\Lambda}$, equivalently $w_{\Lambda}=-1$. Provided enough of the total 
energy density is stored in this component, the expansion will be accelerated. 
A first generalization could be some phenomenological scaling Dark Energy with constant 
equation of state and $w<-\frac{1}{3}$. Some interesting insight can be gained by considering most 
of the universe energy stored in such a component (see e.g. \cite{CPol}).

Another more elaborate alternative to a pure cosmological constant $\Lambda$ is some effective, 
slowling varying, cosmological constant term that will start driving the universe expansion at 
low redshifts. The prominent candidate in this respect is some minimally coupled scalar field 
$\Phi$, often termed quintessence, slowly rolling down its potential so that it has a negative 
pressure, $p_{\Phi}=w_{\Phi}\rho_{\Phi}$, $w_{\Phi}<-\frac{1}{3}$ \cite{RP,Wetterich,FJ,CDS}.
Later on tracking solutions were introduced \cite{ZWS}, a substantial improvement with respect to 
the initial conditions problem, however without solving the cosmic coincidence problem.
Actually, this is precisely the mechanism which drives the inflationary stage in the 
Early Universe and here too, one can consider several potentials and investigate how well they 
fit the observational data. If a scenario of this kind is the correct one, then it is possible in 
principle to reconstruct (the relevant part of) its potential $V(\Phi)$ and the corresponding 
equation of state characterized by $w_{\Phi}$, using luminosity distance measurements in function 
of redshift $z$. This procedure that can be extended to more elaborate models of gravity like 
so-called generalized, or extended, quintessence models in the framework of scalar-tensor theories 
of gravity \cite{BEPS,EP}. Note that the latter case can represent a possible physical realization of 
so-called phantom energy, with $w <- 1$ \cite{BEPS} and it has been intensively investigated 
(see e.g. \cite{PM03}). 
One can also consider other modifications to the theory of gravity or to the coupling of scalar 
fields (see e.g. \cite{GG}).

A Chaplygin gas is another example of more exotic Dark Energy candidates that was proposed, in the 
course of the universe expansion, this gas undergoes a transition from a dust-like to a cosmological 
constant-like equation of state \cite{Chaplygin}.

Actually, as the Dark Energy sector is still unknown, it is interesting to explore 
all possible models that can pass successfully the observational constraints and we want 
to investigate Double Quintessence models for which this sector consists of two coupled 
scalar fields. Two scalar fields models give rise to exchange of kinetic energy between 
the two fields and produce a non-trivial time evolution of the Dark Energy equation of state. 
Some models of two fields quintessence have been introduced and studied in the literature 
\cite{Halyo,Fujii,MPR,BBS}.
We show with simple two-fields models how the introduction of an auxiliary field 
can bring accelerated expansion of our Universe to an end.
Though our models have natural parameters, the cosmological coincidence however is not 
solved here. We note that many-fields inflationary models can have signatures 
\cite{PS95} which distinguish them from single-field inflationary models, in the first 
place the possibility to have a characteristic scale in the primordial fluctuations 
spectrum (see e.g. \cite{PS92}). In this respect, the situation is here more contrasted and we 
will return to this point in the Conclusion.

The existence of an event horizon in the case of a de Sitter phase is an obstacle to the 
implementation of string theory because the S-matrix formulation is no longer possible, hence 
eternal acceleration, leading to a de Sitter space in the asymptotic future is problematic 
\cite{HKS}.  
Therefore, transient acceleration of our universe is certainly a wellcome feature in this respect.
Even more, we will show that we can have, in a fairly natural way, a transient acceleration 
which has already ended at the present time and still in accordance with the observations. 
We explore the window in parameter space for which this intriguing scenario is 
realized. It is even possible, though this is a marginal possibility, to construct scenarios where 
there is no acceleration at all, neither at the present time nor in the past of our Universe. 
Of course, agreement with the observational data still require that the Dark Energy sector dominate 
from some time on and has an effective equation of state satisfying $w<0$. We will see that in our 
models the quantity $w$, which will be called eos (equation of state) parameter, is not 
constant and actually strongly varying on the redshift interval 
$0<z<2$ relevant for (past) luminosity distance measurements. Actually, it was already 
found earlier that luminosity-distance measurements are rather insensitive to large variations of 
the eos parameter at low redshifts \cite{MBS,CPol}.  
This is why, very interestingly, in our Double Quintessence models, it is possible to construct scenarios 
where the standard interpretation of a presently accelerating Universe can be challenged.

In Section II, we give the basic equations of our system and introduce the relevant 
quantities and notations. In Section III, we list the observational constraints on our 
models and the associated quantities. In Section IV, our models are presented and investigated 
numerically in Section V and regions in parameter space leading to viable models are explored. 
We will take $8\pi G=M_p^{-2}=1$ (where $M_p$ is the reduced Planck mass) and 
$\hbar=c=1$ in the following.


\section{Cosmological evolution}

Let us introduce now the background evolution equations of our system. 
The equations of motion for a spatially flat FLRW cosmology with two coupled 
homogeneous scalar fields $\Phi$ and $\Psi$, and Hubble parameter $H$, are 
conveniently written in the following way:
\bea
\dot{\rho_b} &=& -3 H \gamma_b \rho_b \\
\ddot{\Phi} &=& -3H\dot{\Phi} - \partial_\Phi V \\
\ddot{\Psi} &=& -3H\dot{\Psi} - \partial_\Psi V \\
\dot{H} &=& -4\pi G(\gamma_m\rho_m+\gamma_r\rho_r+\gamma_Q\rho_Q)~,
\eea
subject to the constraint equation:
\be
H^2 = \frac{8\pi G}{3}(\rho_m+\rho_r+\rho_Q)-\frac{k}{a^2}~.
\ee
Here a dot denotes a derivative with respect to the cosmic time $t$, the subscript $b$ 
refers to the dominant background quantity, either dust (m) or radiation (r) while $Q$ 
refers to the Dark Energy sector, here the two quintessence scalar fields. 
We have further  
\be
p_b=(\gamma_b-1)\rho_b~, 
\ee
with $\gamma_m=1$ and $\gamma_r=\frac{4}{3}$ where, for any component $i$, we have introduced 
for convenience the quantity 
\be
\gamma_i\equiv 1 + w_i~.
\ee
Finally the quintessence fields with potential $V$ have the following
energy density and pressure:
\bea
\rho_Q =\frac{1}{2}\dot{\Phi}^2 + \frac{1}{2}\dot{\Psi}^2 + V(\Phi,\Psi)\\
p_Q =\frac{1}{2}\dot{\Phi}^2 + \frac{1}{2}\dot{\Psi}^2 - V(\Phi,\Psi)
\eea
with $p_Q = (\gamma_Q-1)\rho_Q$.
It is convenient to define the following new variables :
\be 
X_{\Phi}=\sqrt{\frac{8\pi G}{3H^2}}~\frac{\dot{\Phi}}{\sqrt{2}},~~~
X_{\Psi}=\sqrt{\frac{8\pi G}{3H^2}}~\frac{\dot{\Psi}}{\sqrt{2}},~~~
X_V=\sqrt{\frac{8\pi G}{3H^2}}~\sqrt{V}.
\ee
The above equations are then written in the following way :
\bea
\Phi' &=& (8\pi G)^{-\frac{1}{2}}\sqrt{6}~X_{\Phi}\\
\Psi' &=& (8\pi G)^{-\frac{1}{2}}\sqrt{6}~X_{\Psi}\\
X_{\Phi}' &=& -3X_{\Phi}-\sqrt{\frac{3}{2}}(8\pi G)^{-\frac{1}{2}}~X_V^2~\partial_\Phi{\rm ln}V+\frac{3}{2}~X_{\Phi}F\\
X_{\Psi}' &=& -3X_{\Psi}-\sqrt{\frac{3}{2}}(8\pi G)^{-\frac{1}{2}}~X_V^2~\partial_\Psi{\rm ln}V+\frac{3}{2}~X_{\Psi}F\\
X_V' &=& \sqrt{\frac{3}{2}}(8\pi G)^{-\frac{1}{2}}X_V(X_{\Phi}\partial_\Phi{\rm ln}V+X_{\Psi}\partial_\Psi{\rm ln}V) 
              +\frac{3}{2}~X_VF
\eea
with
\be
F(X_{\Phi},X_{\Psi},X_V,N) = (1-X_{\Phi}^2-X_{\Psi}^2-X_V^2)\left(\frac{\gamma_m\rho_m+\gamma_r\rho_r}{\rho_m+\rho_r}\right)
                             + 2(X_{\Phi}^2+X_{\Psi}^2)
\ee
where a prime denotes a derivative with respect to the quantity $N$, the number of e-folds with 
respect to the present time,
\be
N\equiv {\rm ln} \frac{a}{a_0}~,\lb{N} 
\ee
and we have also $H=\dot{N}$. It is straightforward to relate the quantity $N$ to the redshift 
$z$
\be
1 + z = e^{-N}~.
\ee
The fields are expressed in units of the reduced Planck mass and all quantities above are 
dimensionless.
Moreover we have the following standard evolution for matter and radiation densities
\bea
\rho_m &=& \rho_{m,0}\exp(-3N)~,\\
\rho_r &=& \rho_{r,0}\exp(-4N)~, 
\eea
and more generally for constant $w_X$
\be
\rho_X = \rho_{X,0}\exp(-3[1+w_X]N)~.
\ee
Hereafter and in (\ref{N}), the subscript $0$ refers to the value
of any quantity today and the subscript $i$ refers to its value 
at some initial time $t_i$. 


\section{Observational constraints}

We consider now all the relevant quantities pertaining to our system which will enter the 
observational contraints.

The relative energy density for matter, radiation and quintessence,  $\Omega_m$, $\Omega_r$ and $\Omega_Q$, 
where we have for each component labelled by the subscript $i,~\Omega_i\equiv \frac{\rho_i}{\rho_c}$ with 
$\rho_c$ the critical density, are given by
\bea
\Omega_m &=&(1-X_{\Phi}^2-X_{\Psi}^2-X_V^2)
   \frac{\Omega_{m,0}}{\Omega_{m,0}+\Omega_{r,0}~e^{-N}}~,\\
\Omega_r &=&(1-X_{\Phi}^2-X_{\Psi}^2-X_V^2)
   \frac{\Omega_{r,0}}{\Omega_{r,0}+\Omega_{m,0}~e^N}~,\\
\Omega_Q &=& X_{\Phi}^2+X_{\Psi}^2+X_V^2~.
\eea
The equation of state (eos) parameter $w_Q$ for the Dark Energy (Double Quintessence) sector,
and the effective eos parameter $w_{eff}$ \cite{HWDCS} read
\bea
w_Q &\equiv& \frac{p_Q}{\rho_Q} = \frac{X_{\Phi}^2+X_{\Psi}^2-X_V^2}{X_{\Phi}^2+X_{\Psi}^2+X_V^2}~,\\
w_{eff}&=&\frac{\int_0^{a_0}~da'w_Q(a')\Omega_Q(a')}{\int_0^{a_0}~da'\Omega_Q(a')}
       =\frac{\int_{-\infty}^0~dN'e^{N'}(X_{\Phi}^2+X_{\Psi}^2-X_V^2)}{\int_{-\infty}^0~dN'e^{N'}(X_{\Phi}^2+X_{\Psi}^2+X_V^2)}~.
\eea
The case where Dark Energy consists of one field is straightforwardly recovered from the above equations.  
Finally, the deceleration parameter $q$,
the Hubble-parameter-free luminosity distance $D_L$ and the age of the Universe $t_0$
are respectively given in terms of $N$ by :
\bea
q &\equiv& -\frac{\ddot a}{a H^2} = \frac{1}{2}\sum_i \Omega_i (1 + 3w_i)\lb{q}~,\\   
&=& \frac{1}{2}(1+\Omega_r+3 w_Q\Omega_Q)\lb{q1}~,\\
D_L &\equiv& H_0 d_L = (1+z)\int_0^z~dz' \frac{H_0}{H(z')} = e^{-N}\int_N^{0}~dN' e^{-N'}\frac{H_0}{H}~,\lb{DL}\\
H_0 t_0 &=& \int_{0}^{+\infty}~dz \frac{H_0}{(1+z)H(z)} = \int_{-\infty}^{0}~dN \frac{H_0}{H}~,\lb{t0}
\eea
with 
\be
\left(\frac{H}{H_0}\right)^2 = \frac{\Omega_{m,0}e^{-3N}+\Omega_{r,0}e^{-4N}}
                                    {1-X_{\Phi}^2-X_{\Psi}^2-X_V^2}~.
\ee
Let us consider two universes with the same ``history'', in the sense that they share the quantity 
$h(z)\equiv \frac{H(z)}{H_0}$ while they differ in the present value $H_0$. 
As can be seen immediately from (\ref{DL}), the quantity $D_L$ depends only on $h(z)$ and 
will therefore be the same for both universes. 
It will be interesting to compare luminosity distances $d_L(z)$ for such universes, we have in 
particular when comparing two universes with same $h(z)$ but different Hubble constant $H_0$ 
and $H_{0,\Lambda}$
\be
d_L(z;H_0) = \frac{H_{0,\Lambda}}{H_0} ~d_L(z;H_{0,\Lambda})
           = \frac{h_{\Lambda}}{h} ~d_L(z;H_{0,\Lambda}) \equiv e_{\Lambda} ~d_L(z;H_{0,\Lambda})~,\lb{e1}
\ee
where $h\equiv \frac{H_0}{\rm 100~km/s/Mpc}$, resp. $h_{\Lambda}\equiv \frac{H_{0,\Lambda}}{\rm 100~km/s/Mpc}$. 
We have further for any two universes (dropping the argument $z$)
\be
%
%
\frac{d_L(h)}{d_{L,\Lambda}}(h_{\Lambda}) = e_{\Lambda} \frac{D_L}{D_{L,\Lambda}}  ~,\lb{eD}
\ee
which shows the relative variation of $d_L(z;H_0)$ when varying $H_0$ with respect to some {\it fixed} value 
$H_{0,\Lambda}$. We will make use of (\ref{eD}) in Figure 10.

The traditional sign convention for the ''deceleration'' parameter $q$
gives a positive $q$ for a decelerating universe. 
A cosmological constant $\Lambda$ corresponds to the particular case $w_{\Lambda}=-1$ 
and $\Omega_{\Lambda}=\frac{\Lambda}{3H^2}$. 
For flat universes, the relative energy densities $\Omega_i$ satisfy 
$\sum_i \Omega_i = \Omega_m+\Omega_r+\Omega_Q = 1$ at all times.
In particular, at late times, the energy density of radiation can be neglected and 
(\ref{q},\ref{q1}) gives 
\be
q \simeq \frac{1}{2}\Omega_m + \frac{1}{2}\Omega_Q (1 + 3w_Q)
  \simeq \frac{1}{2}( 1 + 3w_Q~\Omega_Q ) ~.\lb{q2}  
\ee
We see in particular from (\ref{q}) that a decelerated expansion at late times requires
\be
w_Q > -\frac{1}{3}~\Omega_Q^{-1}~.\lb{q3}
\ee
In this work, we will consider universes where the quantity $q$ (re)changes sign, from 
negative to positive, i.e. accelerating universes resuming a decelerated expansion, in some 
cases even before the present time. We will call henceforth $t_{end}$ the time at which the 
transient accelerated stage ends
\be
q(t\geq t_{end})\geq 0~.
\ee 
For our system, the eos parameter $w_Q$ is constrained between $-1<w_Q<1$ while the energy 
density of the Dark Energy sector (like the energy density of any component) is bounded 
according to $0<\Omega_Q<1$. The parameter $q$ has to be negative at the present time if the 
universe is accelerating today, viz. $q_0<0$.

The nucleosynthesis bound is the most stringent one \cite{BHM} 
\be
\Omega_Q(N \sim -23)\lesssim 0.045~~~~~~~~~{\rm at}~2\sigma~,
\ee
while we have at last scattering \cite{BHM}
\be
\Omega_Q(N \sim -8)\lesssim 0.39~~~~~~~~~~~{\rm at}~2\sigma~.
\ee
We adopt the following range $0.6 \lesssim h \lesssim 0.8$ and we consider universes satisfying 
the conservative bounds \cite{SN2,BM,Spergel,CC02,HM02,MMOT03,CMMS}
\bea
0.2 \lesssim &\Omega_{m,0}& \lesssim 0.45~,\lb{OMm}\\
\Omega_{r,0}h^2 &=& 4.3069\times 10^{-5},
\eea
as well as 
\be
0.55 \lesssim \Omega_{Q,0} \lesssim 0.8~.
\lb{Om}
\ee
In \cite{SN2,BM,Spergel}, a constant eos parameter $w_Q$ was considered.
The quantity $w_{eff}$ was introduced in \cite{HWDCS} in order to account for a 
varying equation of state of the Dark Energy sector.
Note that for constant eos parameter one has $w_{eff}=w_Q$ and it was shown in \cite{BM} 
that constraints on $w_Q$ can be replaced by constraints on $w_{eff}$ so we consider 
the observational constraint :
\be
w_{eff} \lesssim -0.70~.\lb{w1}
\ee
The conservative bound (\ref{w1}) is required by the CMB as well as the SNIa 
and the  LSS data \cite{BM,Spergel,CC02,HM02,MMOT03}.

Combining eqs (\ref{q3},\ref{Om}), {\it decelerated} expansion today requires the 
necessary condition
\be
w_{Q,0}\gtrsim -0.606~.\lb{w2}
\ee
We note immediately that conditions (\ref{w1}) and (\ref{w2}) are incompatible 
for a constant equation of state, i.e. $w_{eff}=w_Q={\rm constant}$. 
Both conditions (\ref{w1},\ref{w2}) can be met today when the equation of state 
varies strongly at low redshifts as we will see with concrete models studied 
in this work (see Figures \ref{eosAz},\ref{eosqB}).
The lower bound (\ref{w2}) is always necessary but only sufficient when 
$\Omega_{Q,0}=0.55$. When $\Omega_{Q,0}=0.8$, the condition sufficient for 
decelerated expansion is tighter, namely $w_{Q,0}\geq -\frac{5}{12}\approx -0.416$. 
Clearly, the lower $\Omega_{Q,0}$, the easier it is to implement decelerated 
expansion today.
Finally we impose that the age of the Universe satisfies \cite{aget0}
\be
t_0\gtrsim 13 ~{\rm Gyrs}~,\lb{t00}
\ee
which translates into a constraint on the quantity $H_0$, or equivalently on $h$, as 
can be seen from (\ref{t0}). One has 
\be
H_0^{-1} = 3.0856\times 10^{17} h^{-1}{\rm s} = 9.7776 h^{-1}{\rm Gyrs}~,\lb{H0}
\ee
so that the constraint (\ref{t00}) can be rewritten as
\be
H_0 t_0\gtrsim 1.330~h~.\lb{H0t0}
\ee
Note that in the past, the inclusion of a cosmological constant was invoked 
precisely in order to reconcile an ``old'' universe with a high value for 
$H_0$ while today it is motivated by completely different observations.
The bound (\ref{t00}) is conservative and corresponds to the present 
age of an Einstein-de Sitter universe with $h= 0.65$.
In our simulations we will assume equipartition of the energy
\be
\frac{1}{2} \dot{\Phi}_i^2 \sim \frac{1}{2} \dot{\Psi}_i^2 \sim V_i
\ee
initially at the end of the primordial inflationary stage ($z_i \sim 10^{29}$), but the initial time
could as well be taken at nucleosynthesis or at matter-radiation equality \cite{ML02}.


\section{Models}

We will be interested in universes where the accelerated expansion is brought to an end, or even 
does not take place. 
If the universe is accelerating indefinitely it will exhibit an event horizon such that :
\be
D_H(t') = \int_{t'}^\infty \frac{dt}{a(t)}
\label{DH}
\ee
is finite. In the case of a constant eos parameter $w_Q$ and a quintessence domination
($\Omega_Q \gg \Omega_m,\Omega_r$), we have $a \propto t^{\frac{2}{3(1+w_Q)}}$ and if $w_Q < -1/3$ 
the integral (\ref{DH}) is finite.

It is well-known that the problem of initial conditions for the quintessence 
field can be considerably alleviated by the so-called tracking behaviour whereas the relative density 
$\Omega_Q$ follows the density $\Omega_b$ of the dominating background component, with some specific 
evolution of $w_Q$. We will consider two one-field models and two Double Quintessence models. 

Let us stress first what will be relevant in the framework of either one-field, or  two-fields 
quintessence models with transient acceleration that we study here. 
There are several ways to obtain transient acceleration:

\begin{itemize}

\item the tracking behaviour of the scalar potential is not reached until today, namely fields are 
      frozen in the past with $w_Q \simeq -1,~\Omega_Q\ll 1$, while once the tracking behaviour 
      is reached, eq.(\ref{q3}) must be satisfied. 
      In Section IV.A, we will see that a pure exponential potential can exhibit such a behaviour as 
      studied in \cite{KL,FR,Cline,KK}. Moreover, in \cite{Russo} it is shown that a FLRW universe 
      filled with a scalar field only can undergo a stage of transient acceleration. As we show in Section 
      IV.D, even the quintessence domination can be avoided because of the varying coupling constant 
      of the model studied there.

\item the existence of a minimum or a local flatness in the $\Phi$ direction of the potential can produce 
      an accelerated expansion but the minimum has to diseappear in order to make this acceleration 
      transient. In the model of Albrecht $\&$ Skordis \cite{AS} (section IV.B) for which a feature
      is introduced in a pure exponential potential, the parameters have to be fitted in order for 
      the field to roll over the barrier. In model IV.C, the minimum is dynamical and disappears.
      So in these cases, the actual acceleration takes place outside the scaling regime, which was 
      reached early on in the universe evolution, when the field $\Phi$ approaches his minimum, see 
      section IV.B and IV.C.

\item Still another possibility to obtain transient acceleration is to cancel the scalar potential $V$ :
      this is the case in the hybrid inverse power law potential $V(\Phi,\Psi)=\Psi^2 M^{4+n}\Phi^{-n}$ 
      studied in \cite{Halyo}. On the contrary, the inverse power law potential cannot produce a transient 
      acceleration because $w_Q \rightarrow -1,~\Omega_Q \rightarrow 1$ in the future \cite{RP,ZWS}.
      Other models where $V$ cancels can produce transient acceleration like the oscillatory Dark Energy 
      model with a double exponential potential $V(\Phi)=\left(A\exp(\frac{1}{2}\lambda\Phi)-
      B\exp(-\frac{1}{2}\lambda\Phi)\right)^2$ \cite{RSPC} and also the power-law potential 
      $V\propto \Phi^{2n},~n=1,2,..$.
\end{itemize}

If $V \leq 0$, even a flat universe can undergo a transient acceleration followed by a big crunch 
\cite{ASS,KKLLS}, and an ever decelerating universe can also be considered in a closed FRLW ($k = +1$) 
\cite{Vishwakarma}.
In the following we will restrict ourselves to positive potentials $V \geq 0$ and flat space $k = 0$.
In the transient acceleration picture, it is interesting to consider whether the acceleration 
finishes in a time comparable to $H_0^{-1}$, namely $t_{end} \gtrsim t_0$ or $t_{end} \lesssim t_0$, 
much larger than $H_0^{-1}$, i.e. $t_{end} \gg t_0$, or if the acceleration does not occur at all.

\subsection{Pure exponential potential}

We consider a scalar field $\Phi$ with an exponential potential 
\be
V(\Phi) = M^4 e^{-\lambda \Phi}~, 
\label{expo}
\ee
already widely investigated in the past \cite{Wetterich,FJ,RP,CLW} and motivated in \cite{TN}. 
For this potential an attractor solution exists, either a scaling solution such that :
\be
\Omega_Q = \frac{3\gamma_b}{\lambda^2}~~~~~{\rm and}~~~~~ \gamma_Q = \gamma_b
~~~~~{\rm if}~~~~~ \lambda^2 > 3\gamma_b~,
\label{density1}
\ee
or else a scalar field dominated solution :
\be
\Omega_Q = 1~~~~~{\rm and}~~~~~ \gamma_Q=\frac{\lambda^2}{3} ~~~~~{\rm if}~~~~~ 0 < \lambda^2 < 
                                                                     3\gamma_b~.\label{density2}
\ee
The nucleosynthesis bound $\Omega_Q(1~{\rm MeV})\lesssim 0.045$ at the $2 \sigma$ level 
\cite{BHM} implies from (\ref{density1}) that $\lambda \gtrsim 9$ during the scaling regime.
If we want to take advantage of the attractor property of this potential,
it is impossible to have reasonnable $\Omega_{Q,0}$ without violating the nucleosynthesis bound,
and anyway $w_Q$ would mimic the background component eos, $w_Q = w_m = 0$ thereby preventing a 
past and/or actual acceleration.

In \cite{KL,FR,Cline} the authors present a way to circumvent these arguments and revive
the pure exponential potential, using the attractor property in the future instead of the past 
of the universe by fine tuning the mass scale $M^2 \sim M_p H_0$ (in this Section we will 
sometimes reput $M_p$ for clarity). Contrary to the claim in \cite{KL}, $\dot{\Phi}_i^2 \gg V_i$ 
is not necessary, only $V_i \sim M_p^2 H_0^2$ has to be imposed initially.
Apart from $\lambda$ and $M$, two extra values $\Phi_i$ and $\dot{\Phi}_i$ must be specified. 
We can always take $\Phi_i = 0$ by a redefinition of $M$ which corresponds to a rescaling of 
the problem.

The universe experiences first a kination regime such that
$V \ll \dot{\Phi}^2 \propto a^{-6}$ with $w_Q \simeq 1$
followed by a regime during which $V \gg \dot{\Phi}^2$ and
while $M_p^2 m_{\Phi}^2 \sim V \ll M_p^2 H^2$, the field $\Phi$ gets frozen with $w_Q \simeq -1$ 
until now where $\dot{\Phi}^2 \sim M_p^2 m_{\Phi}^2 \sim V \sim M_p^2 H_0^2$
allowing the attractor regime to be reached in the near future and we are left 
with a fixed ratio, either $\Omega_Q = 1$ and $w_Q = \lambda^2/3 - 1$ if $\lambda^2 < 3$ or else 
$\Omega_Q = 3/\lambda^2$ and $w_Q = 0$ if $\lambda^2 > 3$, in the presence of dust \cite{KL,FR,Cline}.

In contrast to the statement in \cite{FR}, the value $\lambda > \sqrt{3}$ gives a 
viable model provided the scaling regime is not yet reached and it produces 
intriguing possibilities for the evolution of the universe, namely an acceleration which 
ends before today and even {\it no acceleration at all}.

An acceptable quintessence density $\Omega_Q > \Omega_{Q,0} \gtrsim 0.55$ 
implies, using (\ref{density1}), $\lambda \lesssim 2.3$.
If $\lambda < \sqrt{2}$, acceleration is eternal because the attractor solution 
(in the future) is characterized by $\Omega_Q = 1$ and $w_Q <  -1/3$ 
as seen from (\ref{density2}).
Therefore transient acceleration implies $\sqrt{2} < \lambda \lesssim 2.3$ and
an eternal quintessence density domination ($\Omega_Q > \Omega_m$).
Numerical results of section V.A can tighten even more the upper bound.
This model has one fine tuned parameter $M$, which has to satisfy  $M^2 \sim M_p H_0$.

It is possible to produce analogous scenarios containing multiple scalar fields 
with an exponential potential \cite{assisted} of the type 
\be
V = M^4 \sum_{i=1}^{n} e^{-\lambda_i\Phi_i}~.
\ee
In the context of assisted inflation, though each field is unable to support separately 
an inflationnary stage, together they are able to do so. 

But we are interested in a transient acceleration whether already finished or not, hence 
in non inflationnary solutions at late times, which requires the condition
\be
\sum_{i=1}^{n}\frac{1}{\lambda_i^2} < \frac{1}{2}~,
\ee
because both attractor solutions (\ref{density1},\ref{density2}) remain valid with the change
\be
\lambda^2\rightarrow \left(\sum_{i=1}^{n}\frac{1}{\lambda_i^2}\right)^{-1}~.
\ee 
If each $\Phi_i = 0$ initially, then $M$ has to satisfy $M \sim \sqrt{M_p H_0}$ 
with some level of fine tuning. 

Models containing coupled scalar fields with an exponential potential \cite{coupled}
of the type 
\be
V = M^4 e^{-\sum_{j=1}^{m}\lambda_j\Phi_j}~,
\ee
can produce the same scenarios with $\lambda^2$ replaced by $\sum_{j=1}^{m} \lambda_j^2$
in the expressions (\ref{density1},\ref{density2}). So if
\be
2 < \sum_{j=1}^{m}\lambda_j^2~,
\ee
transient acceleration will occur even though for each slope separately $\lambda_j < \sqrt{2}$. 
Again if each $\Phi_j = 0$ initially, then $M$ has to satisfy $M \sim \sqrt{M_pH_0}$.

It is possible to generalize the two last cases \cite{generalized} with the potential
\be
V = M^4 \sum_{i=1}^{n} e^{-\sum_{j=1}^{m_i}\lambda_{ij}\Phi_{ij}}~.
\ee
It can be shown that in the presence of a barotropic fluid, the two late-time attractor
solutions are, either a scaling solution :
\be
\Omega_Q = \frac{3\gamma_b}{\lambda_r^2}~~~~~{\rm and}~~~~~ \gamma_Q = \gamma_b
~~~~~{\rm if}~~~~~ \lambda_r^2 > 3\gamma_b~,
\ee
or else a scalar field dominated solution :
\be
\Omega_Q = 1~~~~~{\rm and}~~~~~ \gamma_Q=\frac{\lambda_r^2}{3} ~~~~~{\rm if}~~~~~ 0 < \lambda_r^2 < 3\gamma_b~.
\ee
with
\be
\frac{1}{\lambda_{r}^2} \equiv \sum_{i=1}^{n}\frac{1}{\sum_{j=1}^{m_i}~\lambda_{ij}^2}~.
\ee
For example if each $\lambda_{ij}=\lambda$ and $m_i = m$ then $\lambda_r = \sqrt{\frac{m}{n}}\lambda$. Thus
assisted inflation tends to lower $\lambda_r$ and the coupled part to increase it, but when
\be
2 < \frac{1}{\sum_{i=1}^{n}\frac{1}{\sum_{j=1}^{m_i}~\lambda_{ij}^2}}~,
\ee
the late-time solution produces a decelerated expansion with the fine tuning $M\sim \sqrt{M_p H_0}$ if 
initially each $\Phi_{ij}=0$.

\subsection{Albrecht $\&$ Skordis potential}

Albrecht and Skordis \cite{AS} proposed an interesting model of quintessence. As noted by \cite{Barrow}, 
this model contains solutions for which there is a transient acceleration of our universe.
We would like to emphasize even more that acceleration which has already ended by today is also a 
possibility.
The model (denoted AS in the following) has the following potential :
\be
V(\Phi) = M^4 e^{-\lambda \Phi}(P_0+(\Phi-\Phi_c)^2).
\label{as}
\ee
The potential eq.(\ref{as}) has a small minimum in order for the field to be trapped at
$\Phi_{\pm} = \Phi_c + (1\pm \sqrt{1-\lambda^2 P_0})/\lambda$.
If $\lambda^2 P_0 < 1$ the minimum exists and the field plays for a while the role of a 
quasi-cosmological constant term ($\Phi$ slows down but doesn't oscillate), 
however if the field has enough kinetic energy it can roll over the barrier.
If $\lambda^2 P_0 > 1$ there is no minimum and the potential has to be flattened 
sufficiently for acceleration to occur and $\Omega_Q$ can reach $0.55$.
In section V.B an accurate interval will be given for $\lambda^2 P_0$.
The only fine tuning in this model is the value of $\Phi_c$ which expresses the cosmological
coincidence problem :
$\Phi_c$ is roughly the minimum of the potential and so defines the moment when acceleration begins.
Once $\Phi_c$ is fixed, the beginning of the accelerated stage is given.
Apart from that, all parameters ($M$,$\lambda$,$P_0$,$\Phi_c$) take natural values.

To summarize, we have to take $\lambda \gtrsim 9$ in order to satisfy the nucleosynthesis bound, 
$\lambda^2 P_0 \sim 1$ in order for the acceleration to be transient and possibly ending before 
the present time, and finally $\Phi_c$ has to be fine tuned so that quintessence dominates today.

\subsection{Pseudo exponential potential}

For the AS potential eq.(\ref{as}) the minimum, when it exists, is fixed. Hence the acceleration 
will be eternal for most parameter values, see Figure \ref{LP}.
With a straightforward generalisation, allowing a simple coupling between two scalar fields, 
we can obtain a transient acceleration, possibly ending before today, whenever acceleration 
takes place.  
The introduction of an auxiliary field $\Psi$ will control the presence or not of a minimum 
for $\Phi$.
The idea is to have initially on one hand a minimum of the potential in the $\Phi$ direction 
responsible for the acceleration of our universe, and on the other hand an evolution of the 
$\Psi$ field such that this minimum disappears in the course of time allowing the resumption of 
matter domination.
We will study a two-fields potential of the form :
\be
V(\Phi,\Psi) = M^4 e^{-\lambda \Phi}(P_0+f(\Psi)(\Phi-\Phi_c)^2+g(\Psi)).
\label{Va}
\ee
The AS model \cite{AS} is recovered when $f \equiv 1$ and $g \equiv 0$.
For the potential eq.(\ref{Va}) the minimum is now located at 
\be
\Phi_{\pm}=\Phi_c+\frac{1}{\lambda}\left(1\pm\sqrt{1-\lambda^2\frac{P_0+g(\Psi)}{f(\Psi)}}\right).
\label{min}
\ee
The function $g$ ($g > 0$) can describe a mass term for $\Psi$ of the form $g \propto \Psi^2$, but it 
is not essential for the dynamics of the model. Thus, we will take $g \equiv 0$ for simplicity. 

The minimum (\ref{min}) will disappear provided we have 
\be
f(\Psi) < \lambda^2 P_0 \equiv f(\Psi_c).
\label{cond}
\ee
We note in passing that the potential can be rewritten in the form
$V(\Phi,\Psi) = M^4/\lambda^2 e^{-\lambda \Phi}(f(\Psi_c)+f(\Psi)(\lambda\Phi-\lambda\Phi_c)^2)$.
We will use for $f$ a positive, continuous function which is monotonic in the region $\Psi > 0$ 
and/or $\Psi < 0$. As can be seen from the condition (\ref{cond}), if $f$ is decreasing, resp. 
increasing, then for $\Psi_i$ smaller, resp. larger, than $\Psi_c$ acceleration is possible 
because the minimum (\ref{min}) exists.
%
%

During the evolution of the universe, the field  $\Phi$ rolls down its potential, which is dominated by the 
exponential part, such that $M_p^2 m^2_{\Phi} \sim M_p^2 m^2_{\Psi} \sim V \sim \dot{\Phi}^2 \sim M_p^2 H^2$
with $\Omega_Q = 4/\lambda^2$, $w_Q = 1/3$ during radiation domination, while $\Omega_Q = 3/\lambda^2$,
$w_Q = 0$ during matter domination until $\Phi$ approaches its minimum. As long as the field $\Phi$ 
is trapped at its minimum (\ref{min}) and oscillates around it, the universe undergoes an accelerated 
expansion 
with $V \gg \dot{\Phi}^2$ and therefore $w_Q \simeq -1$ until $\Psi$ satisfies $\Psi \leq \Psi_c$, 
resp. $\Psi\geq \Psi_c$, if $f$ is increasing, resp. decreasing.
At that moment the minimum (\ref{min}) disappears, allowing $\Phi$ to continue to roll freely
towards larger values and hence the matter dominated regime is resumed.
For a given set of parameters ($\lambda,P_0, ...$), the further the initial condition $\Psi_i$
from $\Psi_c$, the longer the evolution of $\Psi$ toward $\Psi_c$ and so the longer 
the accelerated regime of our universe.
When $\Psi$ is initially larger (smaller) than $\Psi_c$, provided $f$ is increasing (decreasing) 
$\Psi$ passes through $\Psi_c$ because $\partial_\Psi V = M^4e^{-\lambda \Phi}(\Phi-\Phi_c)^2\frac{df}{d\Psi}$ 
is positive (negative), acceleration occurs which is always transient; 
if on the contrary $\Psi_i\leq \Psi_c$ ($\Psi_i\geq \Psi_c$) quintessence domination is not possible.
The critical value $\Psi_c$ controls the presence or not of the minimum for $\Phi$.
Contrary to \cite{Fujii,MPR} where acceleration is always permanent, or \cite{BBS} where 
acceleration can be either transient or permanent, acceleration here is {\it always} 
transient as in \cite{Halyo,RSPC}. Of course the absence of acceleration is also possible in 
principle for all these potentials but this possibility has to be rejected by observations.

For the numerical computations presented in section V.C  we will use a very simple function $f$ without 
any additional parameter, namely
\be
f(\Psi) = \Psi^2~,
\label{f}
\ee
for which the minimum of $\Phi$ diseapears if $-\Psi_c \leq \Psi \leq \Psi_c$ with $\Psi_c \equiv \lambda\sqrt{P_0}$.
Analogously to the AS model, we have to take $\lambda \gtrsim 9$ in order to satisfy the nucleosynthesis 
bound and, once $\Phi_i$ is fixed, $\Phi_c$ has to be fine tuned in order for quintessence domination 
to occur today.

\subsection{Pure exponential potential with a varying coupling constant}

Starting from a pure exponential potential $V(\Phi) \propto e^{-\lambda\Phi}$, we now allow
$\lambda$ to depend on the auxiliary field $\Psi$ and to vary in time, 
i.e. we make the generalization $\lambda \rightarrow \lambda_{eff}(\Psi)$, 
and we consider the potential 
\be
V(\Phi,\Psi) = M^4 e^{-\lambda_{eff}(\Psi)\Phi}~.
\label{Vb}
\ee
A similar idea was used in \cite{AS} where $\lambda_{eff}(\Phi)$ depends on $\Phi$ in order 
to create a minimum and to produce an eternal acceleration. 

We will assume that $\lambda_{eff}$ has a global minimum equal to $\lambda$ in $\Psi = \Psi_{min} = 0$.
Whatever the precise form of $\lambda_{eff}$, 
$\partial_\Phi{\rm ln}V = -\lambda_{eff} < 0$ for all $\Psi$ which implies 
that $\Phi$ is always growing and $\partial_\Psi{\rm ln}V = -\Phi\frac{d\lambda_{eff}}{d\Psi}$ 
is positive, resp. negative, if $\Phi$ and $\Psi$ have opposite signs, resp. the same sign. 
Hence if initially $\Phi_i < 0$ and $\Psi_i > 0$, resp. $\Psi_i<0$, 
(we could take $\Phi_i > 0$ as well with $\lambda_{eff}$ $\rightarrow$ $-\lambda_{eff}$) 
$\Phi$ grows towards positive values 
while $\Psi$ decreases, resp. grows, until it reaches, and oscillates around, zero.
While $\Phi < 0$ and $\Psi \sim 0$, $\lambda_{eff} \rightarrow \lambda$ and $V \rightarrow e^{-\lambda\Phi}$.
Hence, as for the pure exponential $\lambda \lesssim 2.3$ implies sufficient quintessence domination 
($\Omega_{Q,0}\gtrsim 0.55$) after the matter dominated stage.
Once $\Phi > 0$, $\Psi$ moves away from zero and the function $\lambda_{eff}$ is again growing,
allowing for the resumption of matter domination ($\Omega_Q \ll 1$).
The current quintessence domination begins when $\Psi$ goes to zero and consequently when $\lambda_{eff}$ 
reaches its minimum.
Thus $\partial_\Phi{\rm V}$ will only depend on $\lambda$ and so will the dynamics of $\Phi$.
The lower $\lambda$, the slower the evolution of $\Phi$ toward zero and hence
the longer the acceleration and/or the quintessence domination regime of our universe.
Clearly the more remote $\Phi_i$ from $0$, the stronger this effect.  

The quantity $M$ is solely determined by the conditions to have a realistic evolution of 
our universe and turns out to be roughly equal to the energy scale today 
$M \sim \sqrt{M_p H_0}$. In this potential $|\lambda_{eff}\Phi| \leq -\ln M^4/3H^2$ 
because $X_V^2 \leq 1$. So the potential (\ref{Vb}) allows us to use the scaling 
 property of the exponential potential as early 
as the end of inflation without necessarily having kination followed by a stage where 
the fields are frozen. If $X^2_{V,i} \ll 1$, then after a kination regime ($w_Q \sim 1$), 
we have $\rho_Q \ll \rho_b$ and the evolution of the fields is frozen since 
$m^2_{\Phi} \sim m^2_{\Psi} \sim M_p^{-2}V \ll H^2$ until $V \sim M_p^2 H^2$. 
For potential (\ref{Vb}), as for the potential (\ref{Va}), acceleration if it takes place 
is necessarily transient. 
Moreover the full range $0 < \lambda \lesssim 2.3$ gives viable models, either with transient 
acceleration or without acceleration because for this potential the domination of quintessence 
is transient.

We will take for numerical simulations in section V.D :
\be
\lambda_{eff}(\Psi) = \lambda~(1+\alpha\Psi^2)~,\label{Leff}
\ee
with $\alpha > 0$.
To summarize, for the potential (\ref{Vb}), $\Psi$ controls the beginning of quintessence 
domination while $\Phi$ controls its end.
Observational constraints require the following two conditions :
$\lambda \lesssim 2.3$ and $M^2 \sim M_p H_0$.

\section{Numerical results}

We summarize briefly the observational constraints and the initial conditions used 
in our numerical calculations. We will use the following conservative constraints 
(\ref{Om},\ref{w1},\ref{H0t0}) on the quintessence density today $\Omega_{Q,0}$, 
the effective eos parameter $w_{eff}$ and the age of the universe $t_0$:  
\be
0.55\leq \Omega_{Q,0}\leq 0.80, ~~~~~w_{eff}\leq -0.70, ~~~~~H_0 t_0\geq 1.330~h~\lb{constraints}.
\ee

From the equipartition of energy, natural initial conditions suggest that $\Omega_{Q,i} 
\sim 10^{-3}-10^{-4}$ \cite{ZWS}.
When $\Omega_{Q,i} \lesssim 1 ~\%$, the parameter window allowed by observations is unchanged 
and so we will take 
\be
X_{\Phi,i} = X_{\Psi,i} = X_{V,i} = 10^{-2}~~~~~~~~{\rm at}~N_i=-67~,
\ee
implying $\Omega_{Q,i} = 3\times 10^{-4}$ except for the pure exponential potential, Section V.A, 
where we take $X_{\Phi,i}^2 = X_{\Psi,i}^2 = 10^{-4}$ and $X_{V,i}^2$ has to be adjusted.

\subsection{Pure exponential potential}

The exponential potential eq.(\ref{expo}) is a viable candidate for quintessence provided the attractor 
solution is not yet reached as explained in Section IV.A, and in particular 
it can produce transient acceleration.
In addition to the constraint $\lambda > \sqrt{2}$ in order to have transient acceleration 
(see section IV.A),
imposing the observational constraints (\ref{constraints}) only $\lambda \leq 1.975$ is allowed.
In \cite{FR}, only the case $\lambda < \sqrt{3}$ was considered because they used the constraints
$w_{Q,0} \leq -0.60$ and $\Omega_{Q,0} \geq 0.60$ which lead to a stronger constraint on $\lambda$.

Thus in order to have transient acceleration, $\lambda$ is constrained as $\sqrt{2} < \lambda \leq 1.975$
resulting in $w_{eff} \gtrsim -0.86$. Here $3.7 \lesssim X^2_{V,i}\times 10^{113} \lesssim 9.2$ with $\Phi_i=0$
at $N_i=-67$.

Different possibilities arise :
if $(1.82 - 1.837 )\leq \lambda \leq 1.837$, the acceleration ends before today with 
a low quintessence density $\Omega_{Q,0} \lesssim 0.62$ and $w_{eff} \gtrsim -0.75$.
Surprisingly there is {\it no acceleration at all} in the interval $1.838 \leq \lambda \leq 1.975$
while $\Omega_{Q,0} \lesssim 0.61$ and $w_{eff} \gtrsim -0.75$.
If the observations would constrain $w_{eff}$ so that $w_{eff} \lesssim -0.86$, 
only permanent acceleration would be possible (i.e. $\lambda < \sqrt{2}$).

\begin{figure}[h]
\begin{center} 
\psfrag{L}[][][2.0]{$\lambda$}
\psfrag{t}[][][2.0]{$t_{end}/t_0$}
\includegraphics[angle=-90,width=0.75\textwidth]{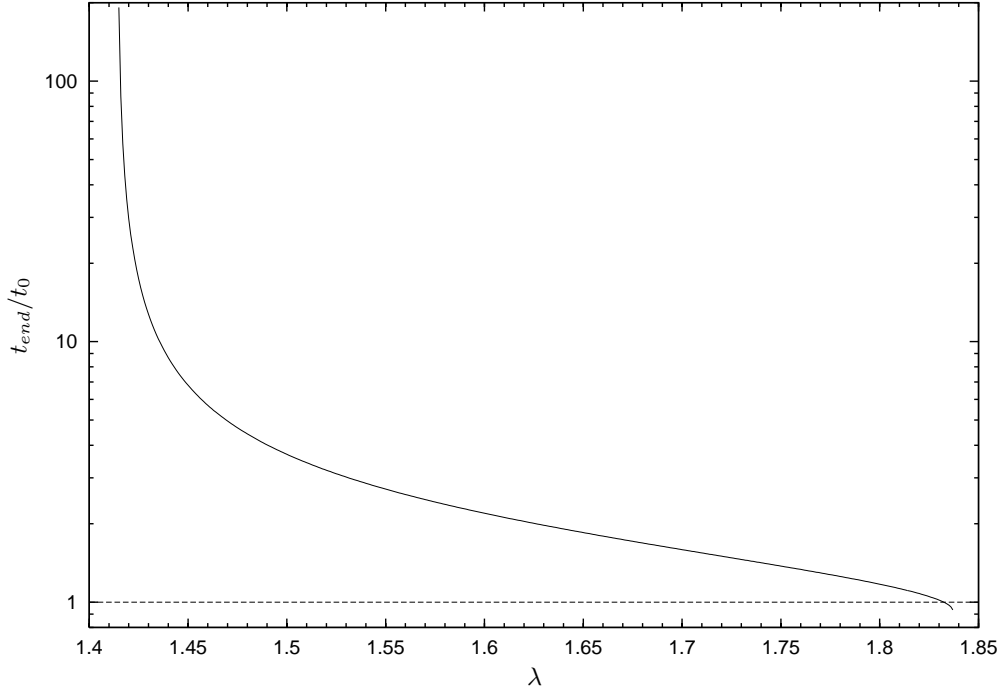}
\end{center}
\caption[]{The ratio $\frac{t_{end}}{t_0}$ is displayed as a function of $\lambda$ 
for the pure exponential potential eq.(\ref{expo}). For $t>t_{end}$, decelerated 
expansion is recovered ($q > 0$). All models displayed have a transient acceleration, 
for those below the dashed line the acceleration is ended by today.   
Initially, at $N_i=-67$, we take $X_{V,i}^2 = V_i/\rho_{c,i} = M^4/\rho_{c,i} = 
5.8\times 10^{-113}$. For this initial value $X_{V,i}$, the range $1.833 \leq \lambda \leq 1.837$ 
yields an accelerated stage which ends before today while the range $1.838 \leq \lambda \leq 1.975$ 
produces {\it no acceleration at all}, the range $\lambda > 1.975$ being excluded by observations, 
eq.(\ref{constraints}).  
Note that $t_{end} \rightarrow + \infty$ when $\lambda \rightarrow \sqrt{2}$.} 
\label{timeB}
\end{figure}

In Fig.~\ref{timeB}, the ratio $t_{end}/t_0$ is shown as a function of $\lambda$.

We should stress that the constraints we have adopted are conservative and a more refined 
analysis would restrain the viability of our models. Let us consider the following constraints 
at 1$\sigma$ from WMAP alone \cite{Spergel}:
\be
13.2~{\rm Gpc}\leq 3~h^{-1}~\int_0^{z_{dec}} \frac{dz}{h(z)}\leq 14.2~{\rm Gpc}~, \lb{IW}
\ee
\be
0.12\lesssim \Omega_{m,0}h^2 \lesssim 0.16~,\lb{OmW}
\ee
\be
0.67\leq h \leq 0.77~,\lb{hW}
\ee
\be
13.1\leq t_0 \leq 13.7~{\rm Gyrs}~.\lb{t0W} 
\ee
We have finally from the SNIa data at 1$\sigma$ \cite{Riess04}
\be
0.26\leq \Omega_{m,0}\leq 0.34~.\lb{OmSN}
\ee
It is easily checked that the range (\ref{OMm}) corresponds to (\ref{OmW}) with the 
lower, resp. upper bound, corresponding to $h\approx 0.81$, resp. $h\approx 0.58$.
However in a more refined analysis the quantities $h$ and $\Omega_{m,0}$ are no longer 
independent.

In Fig.~\ref{dl} the quantity $D_L(z)$ is plotted for an ever decelerating universe 
with $\lambda = 1.84$, $X_{V,i}^2 = 5.8\times 10^{-113}$. This model, representative of the 
scenario without acceleration at all, is characterized by the following cosmological 
parameters $\Omega_{Q,0}=0.583,~w_{Q,0}=-0.568,~w_{eff}=-0.722,~H_0t_0=0.843$. With 
a correction factor $1.11\leq e_{\Lambda} \leq 1.18$, the luminosity distances 
agree with the SNIa data, with uncertainties taken at 1$\sigma$. Taking the fiducial value 
$h_{\Lambda}=0.72$, we would get $0.61 \leq h\leq 0.65$.
For this model we have further $\int_0^{z_{dec}} \frac{dz}{h(z)} = 2.668$ so that the constraint 
(\ref{IW}) yields $0.56\leq h \leq 0.61$ and $0.133\leq \Omega_{m,0}h^2\leq 0.153$ and 
an age $13.51\leq t_0 \leq 14.72~{\rm Gyrs}$.
So this extreme scenario is already in marginal agreement with the WMAP data alone due to 
the low value of $h$, a value $h\simeq 0.61$ being slightly below the 2$\sigma$ error from 
the WMAP data alone. 

\begin{figure}[h]
\begin{center} 
\psfrag{L}[][][2.0]{$\lambda$}
\psfrag{P}[][][2.0]{$\lambda^2 P_0$}
\includegraphics[angle=-90,width=.75\textwidth]{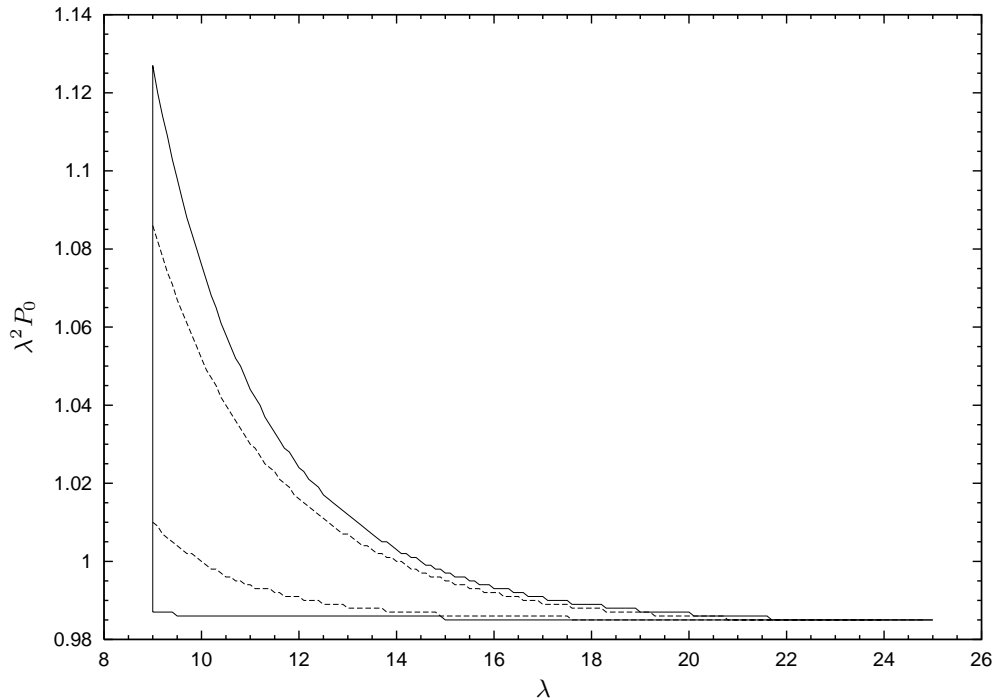}
\end{center}
\caption[]{The allowed window (the area between the solid curves) in the parameter space 
$(\lambda,\lambda^2 P_0)$ is shown for the AS potential eq.(\ref{as}) when the acceleration 
is transient {\it and}  all observational constraints, eq.(\ref{constraints}), are met.
The nucleosynthesis bound implies $\lambda \gtrsim 9$ 
while transient acceleration implies $\lambda \lesssim 25$.
The observations constrain the remaining parameter of the model $\Phi_c$ to be in the range 
$237.5 \lesssim \lambda\Phi_c \lesssim 239$ for the initial condition $\Phi_i = 0$ 
at $N_i = -67$. 
Above the lower dashed line, $q_0$ can be negative as well as positive, above the upper 
dashed line $q_0>0$ always.
For $\lambda^2 P_0$ values below the window, acceleration is eternal as $\Phi$ settles 
at its minimum forever.} 
\label{LP}
\end{figure}

Let us consider now when WMAP is combined with other CMB data and other data probing the power 
spectrum of the perturbations, we get at 1$\sigma$ \cite{Spergel} :

\be
13.7~{\rm Gpc}\leq 3~h^{-1}~\int_0^{z_{dec}} \frac{dz}{h(z)}\leq 14.2~{\rm Gpc}~, \lb{IW+}
\ee
\be
0.126\lesssim \Omega_{m,0}h^2 \lesssim 0.143~,\lb{OmW+}
\ee
\be
0.68\leq h \leq 0.75~,\lb{hW+}
\ee
\be
13.5\leq t_0 \leq 13.9~{\rm Gyrs}~.\lb{t0W+} 
\ee
We get now the tighter bounds $0.56\leq h \leq 0.58$, $0.133\leq \Omega_{m,0}h^2\leq 0.142$.
The value of $h$ is too low and the model without any acceleration is clearly in trouble. 

We should however make the following important remark: the uncertainties are obtained from the data 
assuming a $\Lambda$CDM model and a specific model for the perturbations, a constant spectral index 
$n_s$ for (\ref{IW})-(\ref{t0W}) and a running spectral index for (\ref{IW+})-(\ref{t0W+}). Our models are 
not, by definition, $\Lambda$CDM models so that, strictly speaking, the data should be processed 
specifically for each of our models.

\subsection{Albrecht $\&$ Skordis potential}

\begin{figure}[h]
\begin{center} 
\psfrag{N}[][][2.0]{$N$}
\psfrag{O}[][][2.0]{$\Omega_r,\Omega_m,\Omega_Q$}
\includegraphics[angle=-90,width=.75\textwidth]{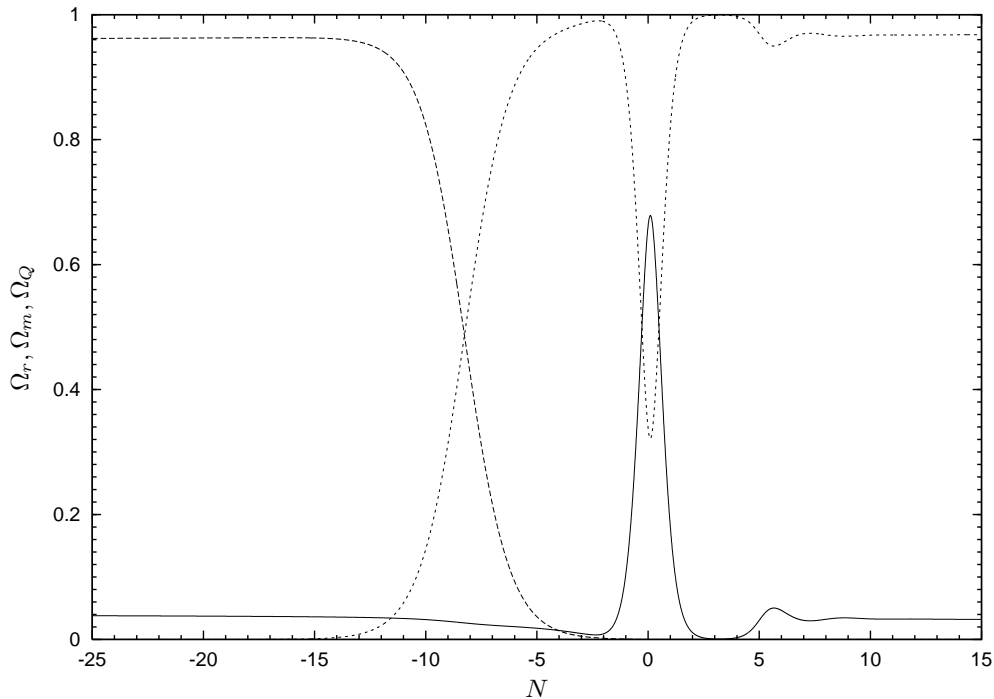}
\end{center}
\caption[]{The evolution of the densities $\Omega_r$ (dotted line), $\Omega_m$ (dashed line) and 
$\Omega_Q$ (solid line) are shown for the potential eq.(\ref{Va'})
with $\lambda = 10$, $P_0 = 0.164$, $\Phi_c = 23.8$ ($\Phi_i = 0$ and $\Psi_i = 5$).
The quintessence density today is $\Omega_{Q,0} \simeq 0.661$. The age of the universe 
is $H_0t_0 \simeq 0.912$ and accelerated expansion stops at $t_{end} \simeq 0.996 ~t_0$. 
The scaling behavior of the quintessence field $\Phi$, eq.(\ref{density1}), is obvious when 
the background is radiation or matter dominated, in the past as well as in the future.}
\label{omegaA}
\end{figure}

\begin{table}[h]
\begin{tabular}{|l|l|l|l|l|l|l|l|}
	\hline
	~$\lambda^2 P_0$~ & ~$\Omega_{Q,0}$~ & ~$w_{eff}$~ & ~$H_0t_0$~ & ~$\frac{t_{end}}{t_0}$~ & ~$w_{Q,0}$~ & ~$q_0$~ \\
	\hline
        1.07 & 0.556 & -0.782 & 0.838 & 0.879 & -0.250 & 0.291 \\
	\hline
        1.06 & 0.573 & -0.804 & 0.849 & 0.907 & -0.303 & 0.240 \\
	\hline
        1.05 & 0.592 & -0.828 & 0.862 & 0.936 & -0.364 & 0.176 \\
	\hline
        1.04 & 0.613 & -0.854 & 0.878 & 0.966 & -0.437 & 0.098 \\
	\hline
        1.031 & 0.634 & -0.879 & 0.895 & 0.997 & -0.516 & 0.009 \\
	\hline
	\hline
	\hline
        1.03 & 0.637 & -0.882 & 0.897 & 1.001 & -0.526 & -0.002 \\
	\hline
        1.02 & 0.663 & -0.913 & 0.920 & 1.042 & -0.634 & -0.131 \\
	\hline
        1.01 & 0.691 & -0.946 & 0.948 & 1.098 & -0.768 & -0.296 \\
	\hline
        1. & 0.719 & -0.977 & 0.978 & 1.194 & -0.914 & -0.486 \\
	\hline
        0.99 & 0.733 & -0.993 & 0.995 & 1.478 & -0.996 & -0.595 \\
	\hline
        0.986 & 0.732 & -0.994 & 0.994 & 2.306 & -1. & -0.598 \\
	\hline
\end{tabular}
\caption[]{Models for the AS potential (\ref{as}) with fixed parameters $\lambda = 10$ 
and $\lambda\Phi_c = 238.5$ are tabulated. All models have transient acceleration, for those 
in the upper part of the Table the expansion is already decelerated today $(t_{end}<t_0)$. 
Note that the model at the top of the lower part of the Table, $\lambda^2 P_0=1.03$, 
satisfies the necessary condition for decelerated expansion today (\ref{w2}), however this 
condition is not sufficient as $\Omega_{Q,0}=0.637$.}
\label{tbl:1}
\end{table}

For the AS potential eq.(\ref{as}), only a tiny interval for $\lambda^2 P_0$ namely 
$0.985 \lesssim \lambda^2 P_0 \lesssim 1.127$
allows a transient acceleration if we impose the constraints (\ref{constraints}), see Fig.~\ref{LP}.
If $\lambda^2 P_0 \lesssim 0.985$, the field $\Phi$ stays at its minimum and acceleration last forever
while if $\lambda^2 P_0 \gtrsim 1.127$ acceleration can occur but $\Omega_{Q,0}$ will never reach the
value $0.55$ because the minimum does not exist and the evolution of the field is too fast.
If the transient acceleration ends before today, again with the constraints (\ref{constraints}) 
the parameter space ($\lambda$, $\lambda^2 P_0$) is even more restricted, see Figure \ref{LP}.
The parameter $\lambda$ is constrained to be $\lambda \gtrsim 9$ because of the nucleosynthesis bound
if the scaling behavior starts earlier than nucleosynthesis, while transient acceleration implies 
$\lambda \lesssim 25$. For this interval, we have $237.5 \lesssim \lambda\Phi_c \lesssim 239$ 
when we start with $\Phi_i=0$ at $N_i=-67$. Moreover $M \sim 10^{-2}\rho_{c,i}^{1/4}$ if we 
impose $X_{V,i}^2 = 10^{-4}$.

\begin{figure}[h]
\begin{center} 
\psfrag{N}[][][2.0]{$N$}
\psfrag{w}[][][2.0]{$w_Q,q$}
\includegraphics[angle=-90,width=.75\textwidth]{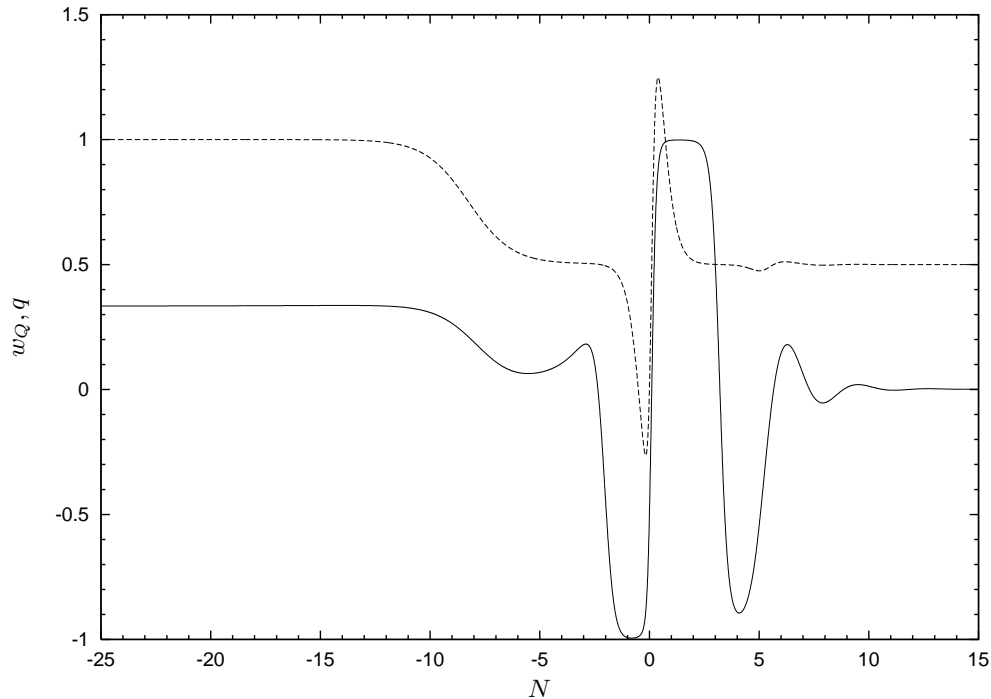}
\end{center}
\caption[]{The evolution of $w_Q$ (solid line) and the deceleration parameter $q$ (dashed line) 
are shown for the model of Figure \ref{omegaA}.
We obtain here $w_{Q,0} \simeq -0.491$, $q_0=0.013$ and $w_{eff} \simeq -0.874$.
Again, we note the scaling behavior of $w_Q$, eq.(\ref{density1}).
In the future we have a kination regime ($w_Q \sim 1$) 
and a freezing regime ($w_Q \sim -1$) of the field $\Phi$.} 
\label{eosqA}
\end{figure}

\begin{table}[h]
\begin{tabular}{|l|l|l|l|l|l|l|l|}
	\hline
	~$P_0$~ & ~$\Omega_{Q,0}$~ & ~$w_{eff}$~ & ~$H_0t_0$~ & ~$\frac{t_{end}}{t_0}$~ & ~$w_{Q,0}$~ & ~$q_0$~ \\
	\hline
        0.17 & 0.575 & -0.771 & 0.846 & 0.889 & -0.207 & 0.321 \\
	\hline
        0.168 & 0.600 & -0.801 & 0.863 & 0.915 & -0.280 & 0.248 \\
	\hline
        0.166 & 0.628 & -0.835 & 0.885 & 0.954 & -0.372 & 0.149 \\
	\hline
        0.164 & 0.661 & -0.874 & 0.912 & 0.996 & -0.491 & 0.013 \\
	\hline
	\hline
	\hline
        0.163 & 0.679 & -0.894 &  0.929 & 1.02 & -0.564 & -0.075 \\
	\hline
        0.162 & 0.699 & -0.916 & 0.949 & 1.05 & -0.649 & -0.181 \\
	\hline
        0.16 & 0.740 & -0.960 & 0.995 & 1.15 & -0.850 & -0.444 \\
 	\hline
        0.15 & 0.746 & -0.989 & 1.008 & 2.49 & -0.993 & -0.611 \\
	\hline
        0.14 & 0.716 & -0.990 & 0.978 & 3.62 & -0.995 & -0.569 \\
	\hline
        0.13 & 0.681 & -0.990 & 0.947 & 5.02 & -0.996 & -0.518 \\
	\hline
        0.12 & 0.643 & -0.989 & 0.917 & 6.80 & -0.998 & -0.463 \\
	\hline
        0.11 & 0.602 & -0.988 & 0.889 & 9.06 & -0.998 & -0.401 \\
	\hline
        0.1  & 0.559 & -0.987 & 0.862 & 11.98 & -0.997 & -0.336 \\
	\hline
\end{tabular}
\caption[]{Models for the potential (\ref{Va'}) with fixed parameters $\lambda = 10$ 
and $\lambda\Phi_c = 238$ are tabulated. The four upper models of the table 
have a transient acceleration already ended by today, see Fig. \ref{eosAz}. 
The model at the top of the lower part of the Table, $P_0=0.163$, 
satisfies the necessary condition for decelerated expansion today (\ref{w2}), however this 
condition is not sufficient as $\Omega_{Q,0}=0.679$.}
\label{tbl:2}
\end{table}

In the table \ref{tbl:1} some examples of a transient acceleration are given with fixed parameters 
$\lambda = 10$, $\lambda\Phi_c = 238.5$ and we vary $P_0$.
All the points for which the transient acceleration is ended by today have $w_{eff} \gtrsim -0.88$.
If $w_{eff}$ is constrained from observations to satisfy $w_{eff} \lesssim -0.88$,
acceleration ended by today is not possible in the AS potential.
Nevertheless transient acceleration ended by today with the AS potential is easier to achieve 
than any transient acceleration using the pure exponential with $w_{eff} \gtrsim -0.86$. 

Whatever $\lambda$ and $P_0$, the maximum value for $t_{end}/t_0$ is $3$ because $P_0$ is bounded 
from below (see also Figure \ref{LP}).

At the end of Section V.C, a model representative of all four models with acceleration ended by 
today will be analysed in the light of the constraints (\ref{IW})-(\ref{t0W+}).

\subsection{Pseudo exponential potential}

\begin{figure}[t]
\begin{center}
\psfrag{R}[][][2.0]{$\Psi_0/\Psi_c$}
\psfrag{t}[][][2.0]{$t_{end}/t_0$}
\includegraphics[angle=-90,width=.75\textwidth]{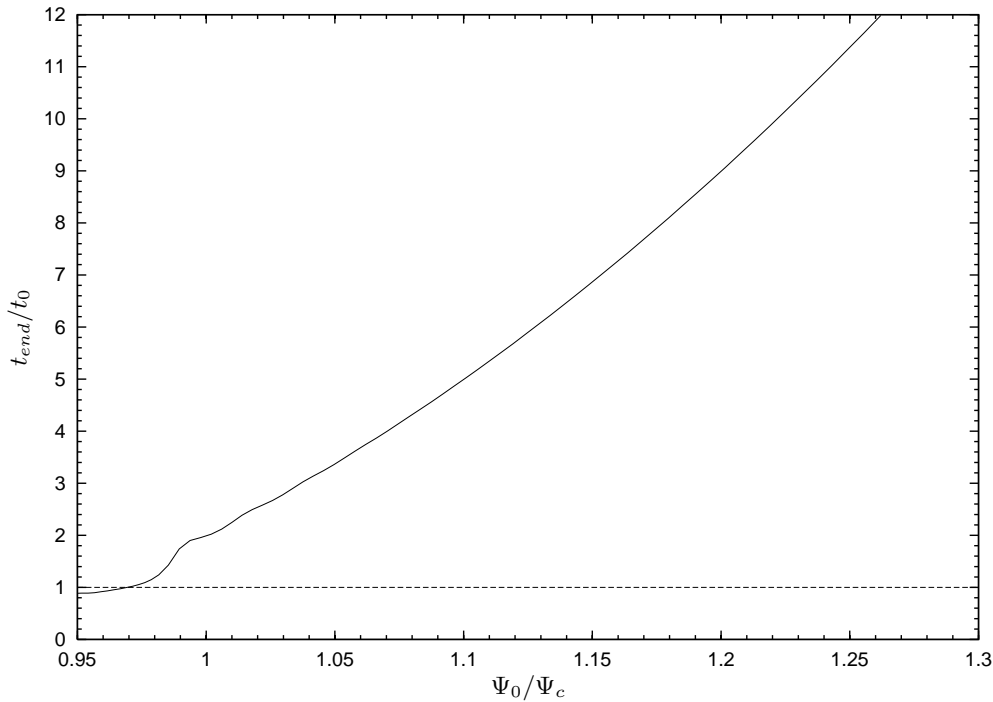}
\end{center}
\caption[]{The ratio $\frac{t_{end}}{t_0}$, where $t_{end}$ is the time at which 
accelerated expansion ends, is shown for the potential eq.(\ref{Va'}) versus the 
ratio $\Psi_0/\Psi_c$ with $\Psi_c = \lambda\sqrt{P_0}$.
The displayed models have the fixed parameters $\lambda = 10$, 
$\Phi_c = 23.8$ with the initial conditions $\Phi_i = 0$ and $\Psi_i = 5$, while 
we vary the parameter $\Psi_c$, or equivalently, $P_0$. All models displayed have transient 
acceleration but only those below the dashed line have a decelerated expansion {\it today}. 
The range of values $\Psi_0/\Psi_c \lesssim 0.95$ is excluded by the observations for this 
set of parameter values.}
\label{timeA}
\end{figure}

We can illustrate the considerations concerning the potential eq.(\ref{Va}) (with $g\equiv 0$) using the 
function defined in eq.(\ref{f}). The potential reads then 
\be
V = \frac{M^4}{\lambda^2} e^{-\lambda \Phi}(\Psi_c^2 + \Psi^2 (\lambda\Phi-\lambda\Phi_c)^2)~,\lb{Va'}
\ee
and the minimum for $\Phi$ disappears if $-\Psi_c \leq \Psi \leq \Psi_c$ with $\Psi_c = \lambda\sqrt{P_0}$.
Here we will use the following initial conditions $\Phi_i = 0$ and $\Psi_i = 5$. Once $\Phi_i$ and 
$\Psi_i$ are given, we can choose $\Psi_c$ so that $0 < \Psi_c < \Psi_i$ while it is 
always possible to find an appropriate $\Phi_c$. In a way analogous to the numerical 
calculations with the AS potential, we keep the parameters $\lambda = 10$, 
$\lambda\Phi_c = 238$ fixed and we vary $\Psi_c$.
Also, as for the AS model, we have $M \sim 10^{-2}\rho_{c,i}^{1/4}$ if we 
impose $X_{V,i}^2 = 10^{-4}$.

In Fig.~\ref{omegaA}, the evolution of densities are plotted for the parameter $P_0 = 0.164$ 
($\Psi_c \simeq 4.05$). This set of parameters induces a transient acceleration which ends before today.
In the scaling regime, $\Omega_Q \simeq 0.04$ during radiation domination and $\Omega_Q \simeq 0.03$ 
during matter domination, while $\Omega_{Q,0} \simeq 0.661$.

In Fig.~\ref{eosqA}, the eos parameter is plotted with $w_{Q,0} \simeq -0.491$ and $w_{eff} = -0.874$ 
and the scaling behavior is evident:
$w_Q \simeq 1/3$ during radiation domination, $w_Q \simeq 0$ during matter domination.
When the field $\Phi$ settles at its minimum, $w_Q \simeq -1$ and this stage is followed by a 
stage of kination and freezing out of $\Phi$ once the minimum is left.
The deceleration parameter $q$ is also plotted implying an actual deceleration as $q_0 \simeq 0.013$.
Acceleration ($q < 0$) begins at $z \simeq 0.658$ and finishes at $z \simeq 0.0035$ when the age
of the universe is $t_{end}/t_0 \simeq 0.996$ while the present age of the universe satisfies 
$H_0t_0 \simeq 0.912$.

\begin{figure}[h]
\begin{center} 
\psfrag{z}[][][2.0]{$z$}
\psfrag{w}[][][2.0]{$w_Q,q$}
\includegraphics[angle=-90,width=.75\textwidth]{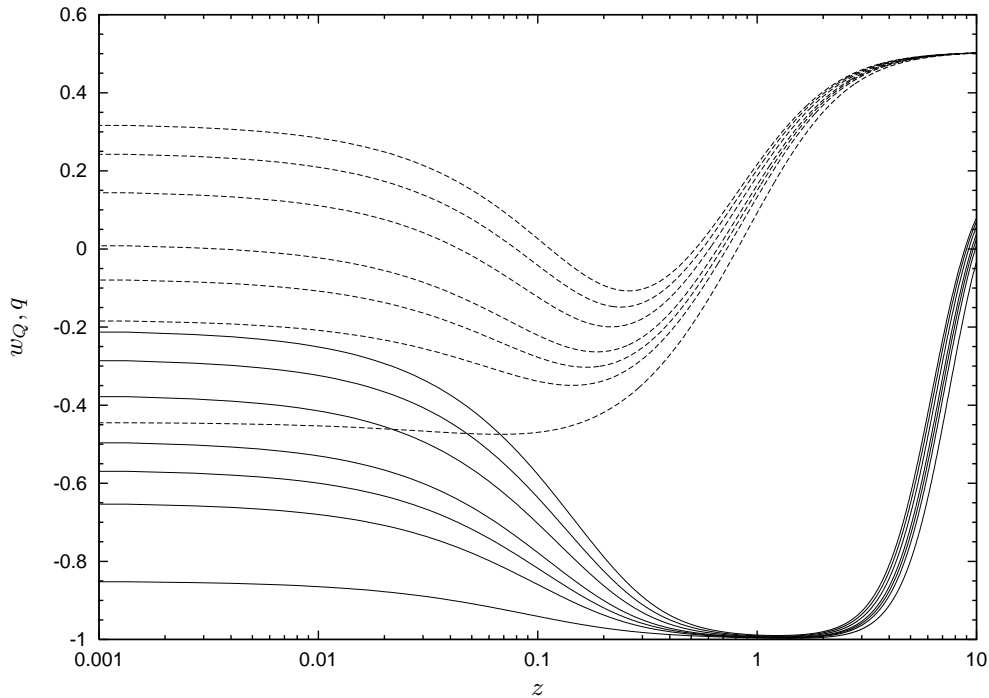}
\end{center}
\caption[]{The evolution of the quantities $w_Q$ (solid line) and $q$ (dashed line) are shown versus 
redshift for the potential eq.(\ref{Va'}) with the fixed parameters $\lambda = 10$, $\Phi_c = 23.8$ while 
$P_0$ takes the values, from bottom to top, $0.160, 0.162, 0.163, 0.164, 0.166, 0.168, 0.170$ 
($\Phi_i = 0$ and $\Psi_i = 5$), see Table \ref{tbl:2}. Note that the third model from bottom, 
$P_0=0.163$, satisfies the necessary condition for decelerated expansion today (\ref{w2}), however 
this is not sufficient because $\Omega_{Q,0}> 0.55$. The models with $P_0 = 0.164, 0.166, 0.168, 0.170$ 
produce a transient acceleration which ends before today ($q_0 > 0$).} 
\label{eosAz}
\end{figure}

In Fig.~\ref{timeA}, the ratio $\frac{t_{end}}{t_0}$ is plotted as a 
function of the ratio $\Psi_0/\Psi_c$. Clearly, the more remote $\Psi_i$ ($\Psi_0$) from $\Psi_c$, 
the longer the acceleration regime. As soon as the minimum disappears, $\Phi$ stops 
oscillating and continues to roll down its potential. Contrary to the AS potential $t_{end}/t_0$ 
can be as large as we want because $P_0$, or equivalently $\Psi_c$, is not bounded from below.

In Fig.~\ref{eosAz}, $w_Q$ and $q$ are plotted against redshift: the cases 
$P_0 = 0.164, 0.166, 0.168, 0.170$ induce an acceleration ended by today.
Viable models where the acceleration ends before today have $w_{eff} \gtrsim -0.88$.

In the table \ref{tbl:2} some examples for this model are shown with 
$\Phi_i = 0$, $\Psi_i = 5$ and again $\lambda = 10$, $\Phi_c = 23.8$ and we 
vary $P_0$ (or equivalently $\Psi_c$).

In Fig.~\ref{dl}, the quantity $D_L$ is plotted for a model for which the acceleration 
is already finished ($\lambda = 10$, $P_0 = 0.164$, $\Phi_c = 23.8$, $\Phi_i = 0$, 
$\Psi_i = 5$, $\Omega_{m,0}=0.339$, $H_0t_0=0.912$). 
With a correction factor $1.03\leq e_{\Lambda} \leq 1.10$, the luminosity distances 
agree with the SNIa data, with uncertainties taken at 1$\sigma$. Taking the fiducial value 
$h_{\Lambda}=0.72$, we would get $0.65 \leq h\leq 0.70$. 

We can repeat for this model a more refined analysis of the kind done for the other model (pure 
exponential) shown in Figures \ref{dl}, \ref{Edl}, see Section V.A. 
We have $\int_0^{z_{dec}} \frac{dz}{h(z)} = 2.992$ so that the constraint 
(\ref{IW}) yields $0.63\leq h \leq 0.68$, $0.135\leq \Omega_{m,0}h^2\leq 0.157$ and an age 
$13.11\leq t_0 \leq 14.11~{\rm Gyrs}$. 
So this model is in good agreement with the WMAP data alone and the SNIa data.
If we consider the tighter constraints (\ref{IW+}), we obtain 
$0.63\leq h \leq 0.66$, $0.135\leq \Omega_{m,0}h^2\leq 0.145$ and an age 
$13.51\leq t_0 \leq 14.11~{\rm Gyrs}$, still allowed by (\ref{IW+})-(\ref{OmW+}).
 
\subsection{Pure exponential potential with a varying coupling constant}

\begin{figure}[t]
\begin{center} 
\psfrag{N}[][][2.0]{$N$}
\psfrag{O}[][][2.0]{$\Omega_Q$}
\includegraphics[angle=-90,width=.75\textwidth]{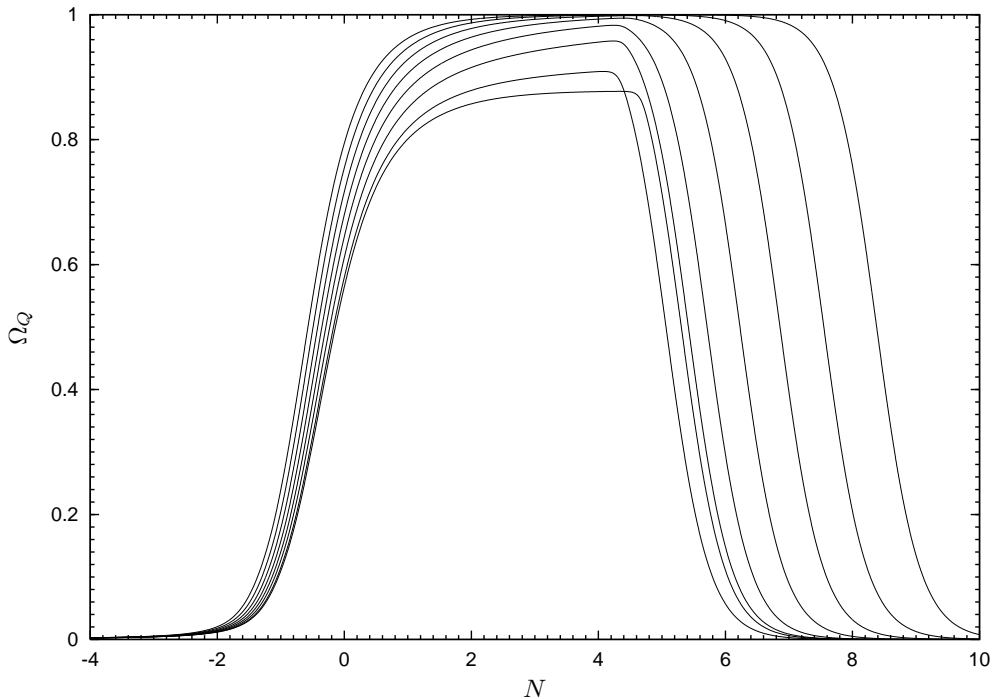}
\end{center}
\caption[]{The evolution of the quintessence density is shown for the potential 
eq.(\ref{Vd}) and same models as in Table \ref{tbl:3})
with, from top to bottom, $\lambda = 1.2, 1.3, 1.4, 1.5, 1.6, 1.7, 1.8, 1.85$.
For the models $\lambda = 1.8, 1.85$ ($\lambda > \sqrt{3}$), there is an intermediate 
stage where $\Omega_Q\to \frac{3}{\lambda^2}\gamma_b$ (cfr eq.(\ref{density1})) 
starting around $N \simeq -1.4$ ($z\simeq 3)$; for the models $\lambda = 1.2, 1.3, 1.4, 1.5, 1.6, 
1.7$ ($\lambda < \sqrt{3}$), there is an intermediate 
stage where $\Omega_Q\to 1$ (cfr eq.(\ref{density2})).
For all models displayed here, $\Omega_Q\to 0$ asymptotically.
Note that for $\alpha = 0$ (pure exponential), the models with $\lambda = 1.2, 1.3, 1.4$ 
($\lambda < \sqrt{2}$, see eq.(\ref{density2})), would yield eternal acceleration 
and domination of the quintessence energy density. 
We see that the smaller $\lambda$, the longer quintessence dominates.} 
\label{omegaB}
\end{figure}

\begin{table}[h]
\begin{tabular}{|l|l|l|l|l|l|l|l|}
	\hline
	~$\lambda$~ & ~$\Omega_{Q,0}$~ & ~$w_{eff}$~ & ~$H_0t_0$~ & ~$\frac{t_{end}}{t_0}$~ & ~$w_{Q,0}$~ & ~$q_0$~ \\
	\hline
	1.2 & 0.794 & -0.806 & 1.016 & 46.42 & -0.715 & -0.352 \\
	\hline
	1.3 & 0.755 & -0.792 & 0.973 & 56.31 & -0.691 & -0.282 \\
	\hline
	1.4 & 0.717 & -0.778 & 0.937 & 63.94 & -0.669 & -0.220 \\
	\hline
	1.5 & 0.680 & -0.765 & 0.907 & 3.62 & -0.649 & -0.161 \\
	\hline
	1.6 & 0.645 & -0.753 & 0.883 & 2.19 & -0.627 & -0.107 \\
	\hline
	1.7 & 0.611 & -0.741 & 0.861 & 1.61 & -0.608 & -0.057 \\
	\hline
	1.8 & 0.580 & -0.730 & 0.843 & 1.20 & -0.590 & -0.013 \\
	\hline
	1.85 & 0.564 & -0.724 & 0.835 & $q>0 ~\forall t$ & -0.582 & 0.007 \\
	\hline
\end{tabular}
\caption[]{Models for the potential eq.(\ref{Vd}) with fixed parameter values 
$\alpha\lambda = 12.44$, and initial conditions $\Phi_i = -5$, $\Psi_i = 2$, 
are shown here. All models have transient acceleration 
except for the model at the bottom of the Table which has no acceleration at all. 
Note that the necessary condition (\ref{w2}) for decelerated expansion today is 
satisfied for $\lambda=1.8$ but it is not sufficient as $\Omega_{Q,0}=0.580$.} 
\label{tbl:3}
\end{table}

We now consider the potential eq.(\ref{Vb},\ref{Leff})
\be
V = M^4~\exp(-\lambda(1+\alpha\Psi^2)\Phi)~,\lb{Vd} 
\ee
for which quintessence domination occurs when $\lambda \lesssim 2.3$ and acceleration today 
takes place provided $M\sim \sqrt{M_p H_0}$.
We have taken $\alpha\lambda = 12.44$ with $\Phi_i = -5$ and $\Psi_i = 2$
and initially $X^2_{V,i} = 10^{-4}$.
Thus the dynamics of the fields starts initially, that is, there is no freezing out ($V \sim H^2$).

In Fig.~\ref{omegaB}, the quintessence density evolution is plotted for different values 
$\lambda = 1.2, 1.3, 1.4, 1.5, 1.6, 1.7, 1.8, 1.85$. 
The effect of the value of $\lambda$ is shown where, for fixed initial 
conditions, the models today are roughly equivalent, but the duration of the
transient acceleration depends strongly on the value of $\lambda$.

\begin{figure}[t]
\begin{center} 
\psfrag{z}[][][2.0]{$z$}
\psfrag{w}[][][2.0]{$w_Q,q$}
\includegraphics[angle=-90,width=.75\textwidth]{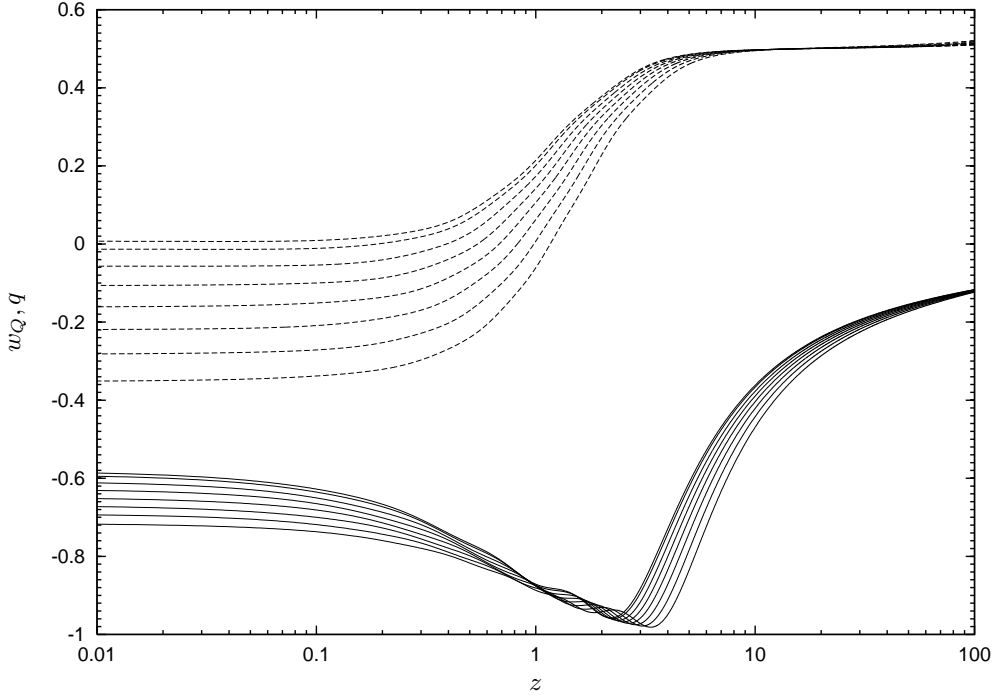}
\end{center}
\caption[]{The evolution of $w_Q$ (solid line) and of the deceleration parameter $q$ 
(dashed line) is shown for the potential eq.(\ref{Vd}) with the same parameters as in 
Figure 7, and with growing values of $\lambda$ from bottom to top.
For the model $\lambda = 1.85$, there is no accelerated expansion at all. 
Note that $w_Q$ exhibits oscillations in the range $z \sim 1-3$ because the auxiliary field 
$\Psi$ reaches, and oscillates around, $0$.}
\label{eosqB}
\end{figure}

\begin{figure}[h]
\begin{center} 
\psfrag{z}[][][2.0]{$z$}
\psfrag{dl}[][][2.0]{$D_L$}
\includegraphics[angle=-90,width=.75\textwidth]{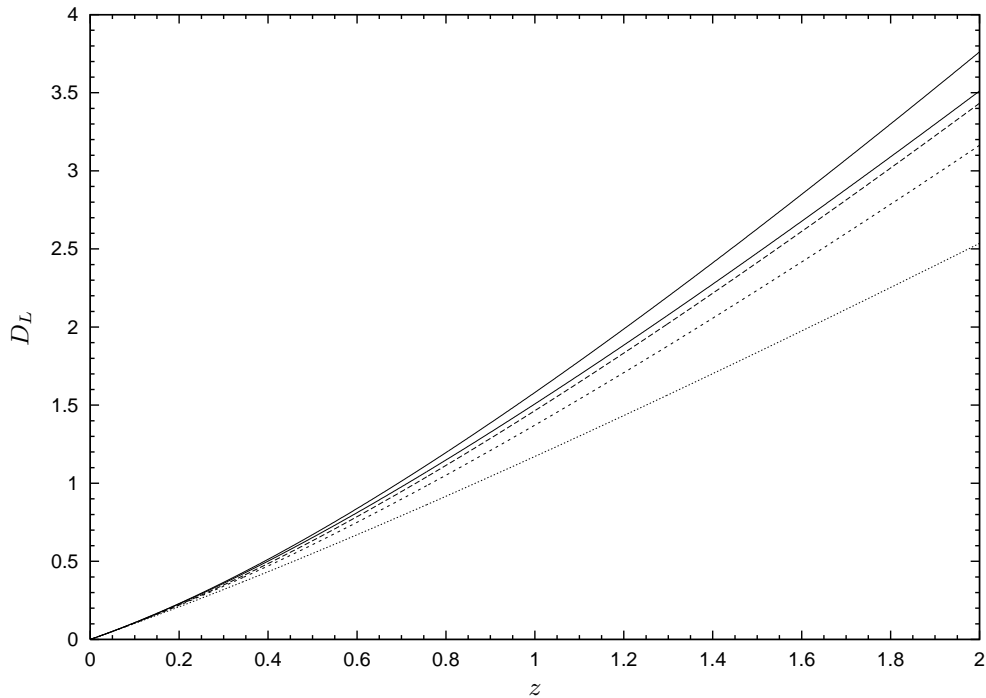}
\end{center}
\caption[]{The following models are plotted: the SNIa data with 1$\sigma$ errors corresponding to flat 
$\Lambda$-models with $\Omega_{\Lambda,0} = 0.74$ (upper solid line), $\Omega_{\Lambda,0} = 0.66$ (lower 
solid line), the Einstein-de Sitter universe (dotted line), 
the potential eq.(\ref{Va'}) for $P_0 = 0.164$ of Table \ref{tbl:2} 
and accelerated expansion ended by today (long-dashed line),   
the pure exponential potential eq.(\ref{expo}) for parameter values $\lambda = 1.84$, 
$X_{V,i}^2 = 5.8\times 10^{-113}$ and {\it no acceleration at all}
(short-dashed line). 
The luminosity distances $d_L(z)$ for a given model with $H_0$ can be compared by varying 
$e_{\Lambda}$, $e_{\Lambda} D_L(z)\equiv \frac{h_{\Lambda}}{h} D_L(z)= H_{0,\Lambda} d_L(z)$, 
where $H_{0,\Lambda}$ (and the corresponding $h_{\Lambda}$) is some fixed fiducial value.
The correction factor $1.03\leq e_{\Lambda}\leq 1.10$ will bring the first model (pseudo exponential) inside the 
SNIa data (1$\sigma$ uncertainties), $1.11\leq e_{\Lambda}\leq 1.18$ is needed for the pure 
exponential.}
\label{dl}
\end{figure}

\begin{figure}[h]
\begin{center} 
\psfrag{z}[][][2.0]{$z$}
\psfrag{Edl}[][][2.0]{$\left(\frac{d_L(h)}{d_{L,\Lambda}(h_{\Lambda})}-1\right) ~\%$}
\includegraphics[angle=-90,width=.75\textwidth]{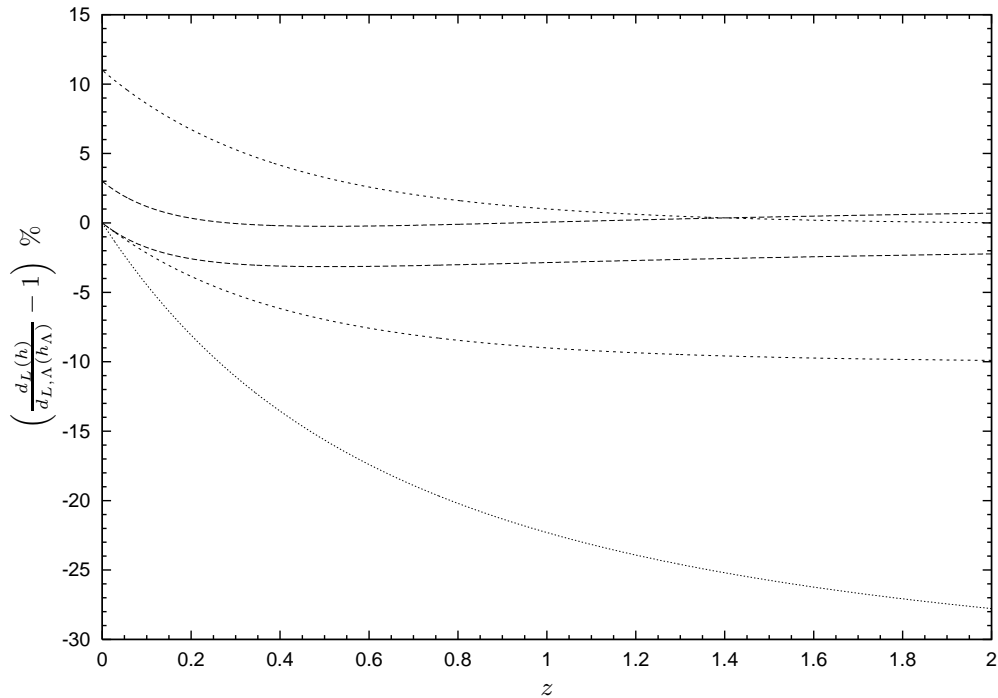}
\end{center}
\caption[]{The relative error for the luminosity distance $d_L(z)$ with respect to the $\Lambda$ model 
with $\Omega_{\Lambda,0} = 0.66$ is shown. One has
$\frac{d_L(h)}{d_{L,\Lambda}(h_{\Lambda})}-1=e_{\Lambda}~\frac{D_L}{D_{L,\Lambda}}-1$. 
The following models are plotted: the Einstein-de Sitter universe with $e_{\Lambda}=1$ (dotted line), 
the potential eq.(\ref{Va'}) for $P_0 = 0.164$ of Table \ref{tbl:2} 
and accelerated expansion ended by today, with $e_{\Lambda}=1$ (lower long-dashed line) 
and $e_{\Lambda}=1.03$ (upper long-dashed line), the pure exponential potential eq.(\ref{expo}) 
for parameter values $\lambda = 1.84$ and $X_{V,i}^2 = 5.8\times 10^{-113}$ 
and {\it no acceleration at all}, with $e_{\Lambda}=1$ (lower short-dashed line) 
and $e_{\Lambda}=1.11$ (upper short-dashed line).
As seen from the Figure, by varying $e_{\Lambda}$, and $e_{\Lambda}>1$, we improve agreement with the 
SNIa data, 
except for small $z\lesssim 0.3$ where the relative departure can be large.
Clearly, this is impossible with the Einstein-de Sitter model.x}
\label{Edl}
\end{figure}

In Fig.~\ref{eosqB}, $w_Q$ and $q$ are shown and we note a permanently decelerating universe 
for $\lambda = 1.85$, however all the models with no acceleration at all have 
$w_{eff} \gtrsim -0.74$ and are marginally viable.
Note that the oscillations of the auxiliary field $\Psi$ around $z \sim 1-3$ translate 
into oscillations in $w_Q$, but it is impossible to see them in the luminosity 
distance $d_L$ \cite{MBS,CPol}.

In the table \ref{tbl:3} below some examples for this model are shown with $\Phi_i = -5$, $\Psi_i = 2$, 
$\alpha\lambda = 12.44$ and we vary $\lambda$.

Scenarios with no acceleration at all in model D are of the same type as for the pure exponential, 
model A. The analysis performed at the end of Section V.A is therefore representative for model D.

\begin{table}[h]
\begin{tabular}{|l|l|l|l|}
	\hline
	~Models~ & ~Scenarios~ & ~Parameters~ & ~Fine tuning~ \\
	\hline
        Pure exponential & eternal acceleration & ~~~~$0 \leq \lambda \leq \sqrt{2}$ & $V_0=M^4e^{-\lambda\Phi_0} \sim M_p^2H_0^2$ \\
	\cline{2-3}
	$V=M^4e^{-\lambda\Phi}$ & transient acceleration & ~~~~$\sqrt{2} < \lambda \lesssim 1.837$ & \\
	\cline{2-3}
	& no acceleration at all & ~~~~$1.838 \lesssim \lambda \lesssim 1.975$ & \\
	\hline
	AS & eternal acceleration & ~~~~$9\lesssim \lambda$ & $\Phi_c$ depends on i.c. \\
	$V=M^4e^{-\lambda\Phi}\left[P_0+(\Phi-\Phi_c)^2\right]$ & & ~~~~$\lambda^2P_0 \lesssim 0.985$ & \\
	\cline{2-3}
        & transient acceleration & ~~~~$9\lesssim \lambda \lesssim 25$ & \\
	& & $0.985 \lesssim \lambda^2P_0 \lesssim 1.127$ & \\
	\hline
	\hline
        AS generalisation & transient acceleration & ~~~~$9\lesssim \lambda$ & $\Phi_c$ depends on i.c. \\
	$V=M^4e^{-\lambda\Phi}\left[P_0+\Psi^2(\Phi-\Phi_c)^2\right]$ & & ~~~~$\lambda^2P_0 \lesssim \Psi_0 \lesssim \Psi_i$ & \\
	\hline
        Exponential with a varying $\lambda$ & transient acceleration & ~~~~$0 < \lambda \lesssim 1.84$ & $M^4 \sim M_p^2H_0^2$ \\
	\cline{2-3}
	$V=M^4e^{-\lambda(1+\alpha\Psi^2)\Phi}$ & no acceleration at all & ~~~~$1.84 \lesssim \lambda \lesssim 2$ & \\
	\hline
\end{tabular}
\caption[]{The possible scenarios for our one-field models (upper part of the Table) and our 
Double Quintessence models (lower part) are summarized.
As can be seen, for the one-field models, eternal acceleration covers most of the allowed parameter space. 
Note that for the Double Quintessence models, the acceleration is necessarily transient.
All four models can produce scenarios satisfying (\ref{constraints}) with a transient acceleration already 
ended at the present time.}
\label{TS}
\end{table}


\section{Conclusion}

Recent observational data suggest that some unknown component, called Dark Energy, contribute 
to about two thirds of the present total energy density filling our universe. Though the 
accuracy of the existing observations allow to constrain already at the present stage to some 
extent the possible Dark Energy candidates, a very large number of models are 
still permitted. It is therefore interesting to investigate all possible scenarios and 
we have also investigated in this work specific models where the Dark Energy sector is made of 
two coupled scalar fields. 
We have studied numerically two one-field models, the pure exponential in 
Section V.A and the AS model in Section V.B, and two Double Quintessence models, which can 
be seen as extensions of the corresponding one-field model, in Section V.C and V.D.
We were interested in particular in scenarios for which the (recent) stage of accelerated   
expansion is transient.

Constraining the models with the observations, we have found the allowed window in the 
corresponding parameter space of each model. Investigation of these models has revealed that 
the following three possibilities can arise: the present 
acceleration is transient and still going on; some accelerated expansion did take place in 
the recent past but is already finished by today; finally no acceleration at all, 
this latter possibility being marginal.  
The two first scenarios, i.e. a transient acceleration either finished or not by today, 
can be obtained in all four models. 
A scenario with no acceleration at all is obtained only for the two one-field models 
V.A and V.D provided the cosmological parameters take their values at the edge 
of the allowed range, $\Omega_{m,0}\gtrsim 0.4$ and $h\lesssim 0.65$.
We summarize the possible dynamics of our models in Table \ref{TS}.

In both Double Quintessence models studied in Sections V.C and V.D, for the allowed window in 
parameter space where accelerated expansion takes place, the dynamics of the model is such 
that the acceleration is necessarily transient, hence for these two models acceleration if it 
takes place will eventually come to an end either before today or in the future. 
In the model of Section V.C, in addition to the tracking field $\Phi$, an auxiliary 
field $\Psi$ is introduced which controls the presence of a minimum for the $\Phi$ 
field and the duration of the Universe transient acceleration. In the model of Section V.D, 
it is the auxiliary field $\Psi$, and hence $\lambda_{eff}$, which induces accelerated expansion 
when it reaches its minimum.
In contrast to our Double Quintessence models where acceleration is necessarily transient, 
for the one-field models we have studied the acceleration is typically eternal and covers 
most of the allowed parameter space.
In view of the theoretical problems posed by eternal acceleration, 
all viable scenarios with transient acceleration constitute a wellcome alternative. 
On the other hand, the cosmological coincidence is not solved here and requires some amount 
of fine tuning on one of the free parameters.
We insist that all the scenarios studied here are in agreement with observations, in particular 
with the Hubble diagram $H(z)$, or the luminosity distances $d_L(z)$ as a function of redshift, 
as reconstructed from the Supernovae data leading to the possible interpretation of a flat universe 
with $\Omega_{\Lambda,0}\simeq 0.72$. 

As mentioned in the Introduction, it is well known that many-fields inflationary models can produce 
primordial fluctiations spectra with a characteristic scale. The question arises naturally whether 
similar effects can be produced here. In fact, the luminosity distances can exhibit a characteristic 
scale if the equation of state of dark energy undergoes a phase-transition. Such cases were considered 
in \cite{BKSU}, where the quantity $w(z)$ was taken with a step-like structure.  
We have checked numerically that such models could indeed exhibit a characteristic scale in their 
luminosity distances. 
In our Double Quintessence models where accelerated expansion is ended at the present time, 
we have also a large variation of $w(z)$ at low redshifts, typically from $w(z)\simeq -1$ up to 
some higher value. However, as can be seen from Figures \ref{eosAz} and \ref{eosqB} this variation 
takes place at very low redshifts $0\leq z\lesssim 0.3$ and, though significant, this variation is 
not sharp enough. This is the reason why no characteristic scale is seen in the corresponding 
luminosity distances.
Actually, it was already emphasized that the luminosity distances are not very sensitive to large, 
smooth, variations of the equation of state, or equivalently of the quantity $w(z)$, at low redshifts $z$. 
Results obtained here are a particular illustration of this property.    

In the light of our results, it is clear that eternal acceleration could be challenged 
in two ways, either the acceleration is transient and will end at some time in the future, 
either it has already ended by today. 
Results obtained here show that the second possibility must be taken seriously if the observations 
allow a rather high matter content today, viz. $\Omega_{m,0}\gtrsim 0.35$ and also 
large variations of the eos parameter $w(z)$ at low redshifts. In particular, if it turns out 
that the data require $w_{Q,0}\lesssim -0.5$, this possibility is ruled out.   

%
%


\end{document}